\documentclass[aps,prd,twocolumn,floatfix,%superscriptaddress,
preprintnumbers,nofootinbib]{revtex4}
\usepackage{amssymb,amsxtra}
\usepackage{graphics}            
\usepackage[dvips]{graphicx} 
\usepackage{sidecap}
\usepackage{latexsym}
\usepackage[T1]{fontenc}
\usepackage[english]{babel}

\def\centeron#1#2{{\setbox0=\hbox{#1}\setbox1=\hbox{#2}\ifdim
    \wd1>\wd0\kern.5\wd1\kern-.5\wd0\fi
    \copy0\kern-.5\wd0\kern-.5\wd1\copy1\ifdim\wd0>\wd1
    \kern.5\wd0\kern-.5\wd1\fi}}

\newcommand{\mc}{\mathcal}

\newcommand{\N}{\mathbb}%Gir ringer, f.eks N

\newcommand{\half}{{\textstyle\frac{1}{2}}}

\newcommand{\be}{\begin{equation}}
\newcommand{\ee}{\end{equation}}
\newcommand{\ba}{\begin{eqnarray}}
\newcommand{\ea}{\end{eqnarray}}
\newcommand{\nn}{\nonumber}
\begin{document}

\title{On the terms violating the custodial symmetry in multi-Higgs-doublet models}
\author{M.~Aa.~Solberg}
\affiliation{Program for Mechanical and Logistic Engineering, HiST, 
N--7004 Trondheim, Norway}
\email{marius.solberg@hist.no}
\affiliation{Department of Physics,  NTNU, N--7491 Trondheim, Norway}

\begin{abstract}{We prove that a generic multi-Higgs-doublet model (NHDM) generally must contain terms in the potential that violate the custodial symmetry. This is done by showing that the $O(4)$ violating terms of the NHDM potential cannot be excluded by imposing a symmetry on the NHDM Lagrangian. Hence we expect higher-order corrections to necessarily introduce such terms.
We also note, in the case of custodially symmetric Higgs-quark couplings, that vacuum alignment will lead to up-down mass degeneration; this is not true if the vacua are not aligned.} 
\end{abstract}
\maketitle

\section{Introduction}
\label{sec:Introduction}
A recent analysis of data from the LHC has proven the existence of a Higgs-like
boson with mass at about $126$ GeV \cite{:2012gk,:2012gu}, while future investigations might give an indication of a sector of scalar particles beyond the single Higgs boson postulated by the standard model (SM). A natural extension of the SM involving
several scalar particles is the SM augmented by several Higgs doublets, resulting in what is 
denoted multi-Higgs-doublet models (NHDM). In these models, some of the Higgs bosons should be
responsible for the generation of
the masses of fermions and the electroweak bosons
\cite{Englert:1964et,Higgs:1964pj,Guralnik:1964eu,Higgs:1966ev}. Other Higgs particles might incorporate the dark matter \cite{Silveira:1985rk,Holz:2001cb,Ma:2006km,Barbieri:2006dq,Patt:2006fw,
Porto:2007ed,Grzadkowski:2009bt,Ivanov:2012hc}. In addition, the NHDM naturally
accommodates CP violation \cite{Lee:1973iz,Weinberg:1976hu,Branco,Nishi:2006tg,Grzadkowski:2009bt,Branco:2011iw}. 

What is usually referred to as the custodial symmetry is an approximate $SO(3)$ symmetry (often denoted $SO(3)_V$) of the SM: 
The symmetry is exact in the limit $g'\to 0$, where $g'$ is the hypercharge coupling constant, in which
the gauge bosons $W^\pm, Z$ form a triplet (with identical masses).
The name "custodial" was chosen since the symmetry guards the tree-level relation 
\ba\label{E:rho}
\rho\equiv \frac{m_W^2}{m_Z^2 \cos \theta_W} =1, 
\ea
from radiative corrections. The radiative corrections are proportional to $g'^2$, and hence vanish in the custodial
symmetric limit $g' \to 0$, for the SM.
What we refer to as "the custodial $SO(4)$ symmetry" 
is the $(SU(2)_L \times SU(2)_R)/\N{Z}_2 \cong SO(4)$ symmetry
which contains the custodial $SO(3)$ symmetry, see sec.~\ref{sec:custodial}. The custodial $SO(4)$ symmetry is, before spontaneous symmetry breaking, exact in the limit $g' \to 0$ for the scalar sector of the SM. 

Adding scalar $SU(2)$ doublets will not change the tree-level relation \eqref{E:rho}. In the limit $g'\to 0$
we get the tree-level relations $m_W^2=m_Z^2=(g^2/4)(|v_1|^2+\ldots +|v_N|^2)$ and $\cos \theta_W=1$.
But additional $O(4)$ violating terms from the potential will contribute to $\Delta \rho$ at loop level. Complex vacuum expectation values (VEVs) will also contribute to $\Delta \rho$ at higher orders of perturbation theory.
If we start with the most general explicitly, i.e.~before spontaneous symmetry breaking, $CP$ conserving NHDM potential, this can be written, after a possible $SU(N)$ scalar basis transformation, as a potential
with arbitrary, but real quadratic and quartic parameters \cite{Gunion:2005ja}. This means that we have no terms linear in the $CP$ violating bilinear operator $\widehat{C}$, cf.~sec.~\ref{sec:BCbid}, while all other parameters in the potential are arbitrary. Hence, the sources of violation of
the custodial $SO(3)$ symmetry of a general, explicitly $CP$ invariant NHDM in the limit $g'\to 0$ are, (i) the $CP$ conserving while $O(4)$ violating terms of the NHDM potential 
(i.e.~terms of the type $\widehat{C}^2$), cf.~sec.~\ref{sec:BCbid}; and (ii) possibly complex VEVs, cf.~\cite{Haber:2010bw} and sec.~\ref{sec:custodial}.
 
Concerning the former source of $\Delta \rho$: The $O(4)$ violating terms of the NHDM potential which are quadratic in the Higgs doublets, will contribute
to $\Delta \rho$ at 1-loop level. Furthermore, the quartic $O(4)$ breaking terms of the NHDM potential will contribute to $\Delta \rho$ at 2-loop level \cite{Grimus:2007if}.\footnote{The only 1-loop corrections from the terms quartic in the Higgs doublets, the "tadpole" diagrams, yield vanishing contributions to $\Delta \rho$ \cite{Grimus:2007if}.}    

Although the $O(4)$ violating terms of the NHDM potential in general do not have to be exactly zero, their magnitude will be constrained by $\Delta \rho$, and also by the oblique parameter $U$. Both $\Delta \rho$ and $U$ are zero in the custodial $SO(3)$ symmetric limit \cite{Peskin:1990zt,Peskin:1991sw}, as we already have seen for $\Delta \rho$. The magnitudes of $\Delta \rho$ and $U$ will grow with the violation of the custodial $SO(3)$ symmetry. The constraints on $\Delta \rho$ and $U$
have for instance already excluded the $CP$ violating 2HDM in a nontrivial region
of the parameter space \cite{Haber:2010bw}. Since an explicitly $CP$ violating 2HDM will violate $CP$ through $O(4)$ violating quadratic terms, the result puts constraints on these $O(4)$ violating terms. Experimental results in the near future may put further restrictions on both the quadratic and quartic $O(4)$ violating terms of the NHDM potential.

In an earlier paper \cite{Olaussen:2010aq} we investigated symmetry properties of the NHDM. 
There we saw that the most general (explicitly) $C$ invariant NHDM potential has an $O(4)$ symmetry (extending the custodial $SO(4)$ symmetry) 
only broken by certain quartic terms in the Higgs fields, of the type $\widehat{C}^2$, cf.~\eqref{E:opNHDM0} below. The transformations $C$ and $CP$ are equivalent for the NHDM, see the discussion following \eqref{E:CovD2}. The terms $\widehat{C}$ are odd under
charge conjugation $C$, and hence terms quadratic in $\widehat{C}$ are invariant under $C$.
 We showed in \cite{Olaussen:2010aq} that, in case of real VEVs, the presence of terms $\lambda^{(3)} \widehat{C}^2$, cf.~\eqref{E:potMHDMCinv}, violates the custodial $SO(3)$ symmetry between the charged and $C$ odd Higgs sectors. We argued that the $C$ odd and charged sectors will get identical mass spectra in the limit $g'\to 0$,
 in case the parameters of the type $\lambda^{(3)}$ initially are set to zero. But when it comes to coupling constant renormalization, terms of the type $\widehat{C}^2$ may show up as counterterms,
 even though their corresponding parameters $\lambda^{(3)}$ initially are set to zero \cite{Olaussen:2010aq}.
   
   Thus a question arises: Is it possible to find a discrete symmetry $D$ which is a symmetry of the $C$ invariant Higgs sector, except that it expels the $O(4)$-violating terms, i.e.~the terms of the type 
 $\widehat{C}^2$, from the NHDM potential when it is imposed on the NHDM Lagrangian? If there exists such a symmetry, we may impose this symmetry on the NHDM Lagrangian, and hence avoid terms of the type $\lambda^{(3)} \widehat{C}^2$ at all momentum scales.
 
 The discrete symmetry $D$ has evidently to lie beyond the symmetry
 group of $\widehat{C}^2$ (which we denoted $P$), but has to be a symmetry of the (other) terms of the Higgs Lagrangian. Hence it has to be an element of $O(4)$, which is the largest possible symmetry group of a Lagrangian which can be expressed by real fields organized in quadruplets, provided that
the NHDM potential is complicated enough so that different Higgs quadruplets cannot be transformed into each other.
As we considered in \cite{Olaussen:2010aq}, the symmetry group of the kinetic Higgs terms may be larger than $O(4)$ if fields 
 with different Higgs indices can be transformed into each other: We there showed that the kinetic
 terms where invariant under $SU(2)\times U(N)$ in the case of $N$ complex Higgs doublets. 
 The $U(N)$ component will generally not be a symmetry of the Higgs potential, since the Higgs fields
 generally will occur in an asymmetric manner in a potential. 
 An element of this $U(N)$ component of the symmetry group of the kinetic terms, that also is a symmetry of a specific NHDM potential, is denoted a Higgs family (HF) symmetry of that potential \cite{Ferreira:2008zy}. In appendix \ref{sec:HiggsFamilySymmetries} we show the only HF symmetry
 of the quadratic (in the Higgs fields) part of the general $C$ (i.e.~$CP$) invariant NHDM potential is an $U(1)$ transformation. This is 
 a symmetry of all terms in a NHDM potential, hence a HF symmetry cannot be imposed to prevent the terms of the type $\widehat{C}^2$ (and these terms only) from occurring in the most general $C$ invariant NHDM potential. 
 
 Hence we can state that a symmetry transformation of a NHDM Lagrangian with a sufficiently complicated potential has to be an element of $O(4)$ containing the custodial $SO(4)$ symmetry. In section \ref{sec:TheCustodialSymmetry} we show that no element
 of $O^-(4)$ (orthogonal matrices with negative determinant) is a symmetry transformation of the kinetic Higgs terms, and hence an exact (discrete or continuous) symmetry of the scalar SM Lagrangian has to be
  an exact subsymmetry of the approximate custodial $SO(4)$ symmetry. In section \ref{sec:TheEffectOfTheAdjointAction} we show that all these exact subsymmetries of the approximate custodial $SO(4)$ symmetry also are symmetries of the terms $ \widehat{C}^2$, and hence there
  is no symmetry that can be imposed on the general $C$ invariant NHDM, which expels the terms of the type 
  $\widehat{C}^2$ only. Equivalently, there is no symmetry that can be imposed on the general NHDM, which expels the $O(4)$ violating terms of the general NHDM only.
	
	In theories with exact extended supersymmetry we may avoid terms of the type $\lambda^{(3)} \widehat{C}^2$. Here we may set $\lambda^{(3)}$
	to zero at one momentum scale, and it will remain zero at all scales, by the non-renormalization theorem \cite{Grisaru:1979wc}. This does not contradict our result, since we are only considering symmetries of the NHDM, 
	i.e.~bosonic symmetries. 
 
This paper is organized as follows: In section \ref{sec:custodial} we revise the custodial symmetry
of the standard model (SM). In section \ref{sec:BCbid} we reinvestigate the symmetry properties of the
operators which are the building blocks of the NHDM potential, in terms of bidoublets---that is the Higgs fields
represented by $4 \times 4$ complex matrices. Moreover we introduce some terminology, and quote some results
from \cite{Olaussen:2010aq}. The subject of section \ref{sec:SymmetriesOfTheKineticTerms} is to what extent the $O(4)$-symmetry containing the custodial $SO(4)$-symmetry is a symmetry of the different parts of the Higgs and gauge field Lagrangian. Some physical consequences of having custodial symmetric Yukawa couplings are discussed in section \ref{sec:Yukawa}. Finally, in section \ref{sec:TheAdjointRepresentationOfO4} we search for symmetries beyond the global
$SU(2)_L\times U(1)_Y$ for the kinetic Higgs terms, by investigating the adjoint action of $O(4)$ on the Lie algebra $u(2)$, or more concretely, the well-defined $O(4)$ transformations of the gauge bosons,
see \eqref{E:mcTO4trafo} below and the discussion thereof. Some mathematical discussions are delegated to the appendices.

\section{The approximate $SO(4)$ symmetry of the SM Higgs Lagrangian}
\label{sec:custodial}
Let the complex SM Higgs doublet be written
\begin{align}\label{E:SMdoub}
	   \Phi= 
 \left(\begin{array}{c}
	\phi^+ \\ \phi^0 
\end{array}\right).
\end{align}
Then the global $SO(4)=\left(SU(2)_L\times SU(2)_R\right)/\N{Z}_2$ custodial symmetry \cite{Sikivie:1980hm} of the SM Higgs Lagrangian (in the limit $g'\to 0$)
 can be made manifest by rewriting the
Higgs doublet as a matrix (bidoublet)
\begin{align}\label{E:MatrixPhi}
	\check{\Phi}=\left(\begin{array}{rr}
	\phi^{0\ast}&\phi^+ \\- \phi^{+\ast}&\phi^0 
\end{array}\right)=\left(\begin{array}{rr}
	v^\ast +\eta - i \phi_3 &\phi_1+i\phi_2 \\- \phi_1+i\phi_2&v+\eta +i \phi_3
\end{array}\right),
\end{align}
  where $v$ is the VEV.
   Here the Higgs potential will be a function of $\text{Tr}[\check{\Phi}^\dag\check{\Phi}]$, and is hence invariant under the global transformation
\begin{align}\label{E:SU2t2}
	\check{\Phi} \to U_L \check{\Phi} U_R^\dag,
\end{align}
where $U_L$ and $U_R$ are $SU(2)$ matrices. The matrix $U_L$ represents the usual gauged $SU(2)_L$ invariance. On the other hand, $U_R$ represents an ordinary global transformation (where the gauge fields, when $g'\to 0$, does not transform in parallel with the Higgs doublet).
 
  The SM scalar Lagrangian density can then be written 
\begin{align}\label{E:LHTr}
	\mc{L}=\frac{1}{2}\left(\text{Tr}[(D_\mu \check{\Phi})^\dag D^\mu \check{\Phi}]+\mu^2 \text{Tr}[\check{\Phi}^\dag\check{\Phi}]-\lambda\text{Tr}[\check{\Phi}^\dag\check{\Phi}\check{\Phi}^\dag\check{\Phi}]\right),
\end{align}
where the covariant derivative in the present notation is
\begin{align}\label{E:CovD}
D_\mu \check{\Phi}=\partial_\mu\check{\Phi}+\frac{1}{2}igW_{i\mu}\sigma_i\check{\Phi}-\frac{1}{2}ig'B_\mu\check{\Phi} \sigma_3,
\end{align}
where $\sigma_i$ are the Pauli matrices. We see that the last term breaks the
$SU(2)_R$ symmetry because of the factor $\sigma_3$. However, in the limit $g'\to 0$ the whole SM scalar Lagrangian has the
full $SU(2)_L\times SU(2)_R$ symmetry, when the $W$ fields transform (as usual) as a triplet
under (the gauged) $SU(2)_L$ and as a singlet under the global $SU(2)_R$ symmetry.
Some authors refer to $SU(2)_R$ as the custodial symmetry \cite{Georgi:1994qn}, others use the term of the symmetry $SU(2)_{L+R}/\N{Z}_2=SO(3)_V$ which
leaves the SM vacuum invariant \cite{Sikivie:1980hm}: 
The VEV of $\check{\Phi}$ can be written
\ba
  \check{\Phi}^0=\left(\begin{array}{rr}
	v^\ast &0 \\0\; &v \end{array}\right),
\ea
and it is invariant under the transformation \eqref{E:SU2t2}, $\check{\Phi}^0\to U_L \check{\Phi}^0 U_R^\dag$, if $U_L=U_R$ and if the VEV $v$ is taken to be real. In the NHDM, complex vacua may violate $SO(3)_V$ in the same manner as in the SM \cite{Haber:2010bw}.
In this article we denote the $SO(4)\cong(SU(2)_L \times SU(2)_R)/\N{Z}_2$ symmetry as the custodial
$SO(4)$ symmetry, as in \cite{Nishi:2011gc}.

\section{The operators $\widehat{B}$ and $\widehat{C}$ of the NHDM in the bidoublet formulation $\check{\Phi}$}
\label{sec:BCbid}

Consider the bilinear (i.e.~linear in both $\Phi_m$ and $\Phi_n$), hermitian operators $\widehat{B}$ and $\widehat{C}$ 
introduced in (2.3) of \cite{Olaussen:2010aq}, defined by\footnote{The operator $\widehat{A}_m$ defined in \cite{Olaussen:2010aq}
equals $\widehat{B}_{mm}$ (no sum over $m$).}
\begin{align}\label{E:opNHDM0}
 \widehat{B}_{mn} &= \widehat{B}(\Phi_m,\Phi_n)= \phantom{-}\half(\Phi_m^\dag \Phi_{n} + \Phi_{n}^\dag
 \Phi_{m}) \nn \\
 \widehat{C}_{mn} &= \widehat{C}(\Phi_m,\Phi_n)=-{\textstyle\frac{i}{2}}
 (\Phi_m^\dag \Phi_{n} - \Phi_{n}^\dag
 \Phi_{m}),
\end{align}
where $\Phi_i$ refers to a doublet of the form \eqref{E:SMdoub}, where every scalar field has an additional index $i$,
\begin{align}\label{E:SMdoub2}
	   \Phi_j= 
 \left(\begin{array}{c}
	\phi^+_j \\ \phi^0_j 
\end{array}\right)=\left(\begin{array}{c}
	\phi_{1j}+i\phi_{2j} \\ v_j+\eta_j +i\chi_j 
\end{array}\right), \quad j=1,\ldots,N.
\end{align}
In \eqref{E:opNHDM0} we let $1\leq m \leq n \leq N$.
The most general potential $V(\Phi_1, \ldots, \Phi_N)$ of the NHDM can then be built up by products and sums of the operators in \eqref{E:opNHDM0}. 
The most general $C$ invariant NHDM potential can then be written
\begin{align}\label{E:potMHDMCinv}
  V_C(\Phi_1, \ldots,\Phi_N) &= 
  \mu_{mn} \widehat{B}_{mn}  
  + \lambda_{mn,m'n'}^{(2)} 
  \widehat{B}_{mn}\widehat{B}_{m'n'} \nonumber \\ &+ \lambda_{mn,m'n'}^{(3)}
  \widehat{C}_{mn}\widehat{C}_{m'n'}, 
\end{align}
  with an implicit sum over repeated indices, where $1\leq m\leq n \leq N$ and $1\leq m'\leq n' \leq N$
  for terms containing $\widehat{B}$, and $1\leq m < n \leq N$ and $1\leq m'< n' \leq N$ for terms
  containing $\widehat{C}$. We also demand that the two pairs of indices $(mn)(m'n')$ are "lexicographically"
  ordered to avoid double counting. This means e.g.~that indices $(12)(13)$ are included, while indices
  $(13)(12)$ are excluded in the sums of \eqref{E:potMHDMCinv}. See sec.~2 in \cite{Olaussen:2010aq} for a more compact indexing.
  Superscripts in the parameters $\lambda$ are chosen to coincide with the notation in  
  \cite{Olaussen:2010aq}. 

The corresponding NHDM Lagrangian density is then given by 
\begin{align}\label{E:fullNHDMpot}
 \mc{L}(x)= &\sum_{m=1}^{N}\; [D^\mu \Phi_m(x)]^\dag [D_\mu \Phi_m(x)] \nn \\
 &- V(\Phi_1, \Phi_2,\ldots, \Phi_N),
\end{align}
where the covariant derivative here is defined as 
\begin{align}\label{E:CovD2}
D^{\mu} = \partial^\mu + ig\frac{\sigma^i}{2}W_i^\mu + ig'YB^\mu.
\end{align}
The term `NHDM Lagrangian' is often used about the Lagrangian density \eqref{E:fullNHDMpot}.
We will apply the term `NHDM Lagrangian' as referring to the Lagrangian density \eqref{E:fullNHDMpot} 
augmented by the (kinetic) gauge field Lagrangian given by  \eqref{E:GFTU0} below.
For the NHDM Lagrangian, charge conjugation $C$ and combined charge conjugation and parity transformations, $CP$, are equivalent. This means the NHDM Lagrangian 
(or more precisely, the action thereof) is $C$ invariant if and only if it is $CP$ invariant. Actually, it is only the NHDM potential that may be explicitly $C$/$CP$ violating, all other terms in the NHDM Lagrangian obey $C$/$CP$ transformations. 
Moreover, spontaneous $CP$ violation is equivalent with spontaneous $C$ violation, as shown in \cite{Olaussen:2010aq}. Hence the 
transformations $C$ and $CP$ are, in essence, interchangeable in this article. We prefer most of the time, for the sake of simplicity, to use
the notion `$C$ invariant' etc, and not to involve the parity transformation $P$: The effect of charge conjugation $C$ on (real representations of) complex Higgs doublets may be implemented by a matrix $C\in SO(4)$ cf.~\eqref{E:matrixC}, while this is not the case for the space-time transformation $P$. Still, if we want to confer our results e.g.~with aspects of the $CP$ (while not $C$) conserving parts of the electroweak Lagrangian, we could replace the term `$C$ invariant' by `$CP$ invariant' and so on.

In the same manner as in the last section, we will now show that the operator $\widehat{C}$ does not share the
$SO(4)$ symmetry held by the rest of the NHDM potential: 
Let $\check{\Phi}_{i}$ refer to a bidoublet of the form \eqref{E:MatrixPhi}, where every scalar field has an additional index $i$.   
A simple calculation shows that
\begin{align}
	\widehat{B}(\Phi_m,\Phi_n)= \frac{1}{2}\text{Tr}(\widehat{B}(\check{\Phi}_m,\check{\Phi}_n)),
\end{align} 
while
\begin{align}
	\widehat{C}(\Phi_m,\Phi_n) &= -\frac{1}{2}\text{Tr}(\widehat{C}(\check{\Phi}_m,\check{\Phi}_n)\sigma_3)\nn \\
	&=
	-\frac{1}{2}\text{Tr}(\sigma_3\widehat{C}(\check{\Phi}_m,\check{\Phi}_n)).
\end{align}
The latter confirms that the operator $\widehat{C}$, in contrast to $\widehat{B}$, does not have
the $SU(2)_R$ [and hence neither the $SO(4)$-] symmetry since the presence of the factor $\sigma_3$ hinders us from  utilizing the cyclic property of the trace. Furthermore, the identity
\begin{align}
	&\widehat{C}(\Phi_m,\Phi_n)\widehat{C}(\Phi_{m'},\Phi_{n'})\nn \\ &= -\frac{1}{2}\text{Tr}(\widehat{C}(\check{\Phi}_m,\check{\Phi}_n) \sigma_3\widehat{C}(\check{\Phi}_{m'},\check{\Phi}_{n'})\sigma_3) \nn \\
	&=
	-\frac{1}{2}\text{Tr}(\sigma_3\widehat{C}(\check{\Phi}_m,\check{\Phi}_n)\sigma_3 \widehat{C}(\check{\Phi}_{m'},\check{\Phi}_{n'})),
\end{align}
infers that operators of the type $\widehat{C}^2$, in the same manner as for $\widehat{C}$ above, do not share the
$SO(4)$ symmetry of a NHDM potential built up by the operators of the types $\widehat{B}$ and $\widehat{B}^2$ only. This is also shown in section 2.2 of \cite{Olaussen:2010aq}, but in a different manner: Here we showed that the symmetry group of $\widehat{C}$ is $Sp(2,\N{R})$, while the
symmetry group of $\widehat{C}^2$ is $P(2,\N{R})$, with 
\begin{align}\label{E:symGrC^2}
	P(2,\N{R}) = Sp(2,\N{R})\cup P(2,\N{R})^-.
\end{align}
Here the real symplectic group $Sp(2,\N{R})$ is defined by
\begin{align}\label{Def:Pk}
	Sp(k,\N{R})=\{ S\in GL_{2k}(\N{R}) | S^T {\cal J} S =  {\cal J} \},
\end{align}
with
\begin{align}\label{Def:Jk}
	{\cal J}=\left(\begin{array}{rr}0_k&I_k\\-I_k&0_k\end{array}\right),
\end{align}
where $0_k$ and $I_k$ is the $k\times k$ zero matrix and the $k\times k$ identity matrix, respectively.
The component $P(2,\N{R})^-$ of the Lie group $P(2,\N{R})$, defined by 
\begin{align}
	P(k,\N{R})^-=\{ S\in GL_{2k}(\N{R}) | S^T {\cal J} S = - {\cal J} \},
\end{align}
  consists of matrices with determinant 
\begin{align}\label{E:Det(-1)^k}
	\det(P(k,\N{R})^-) =(-1)^k.
\end{align}
In the relevant case $k=2$, $P^-$ evidently consists of $4 \times 4$ matrices with determinant $1$.
The same is true for $Sp(2,\N{R})$.

\begin{figure*}
\begin{center}
\includegraphics[width=0.75\textwidth]%[scale=1.1,angle=0]
{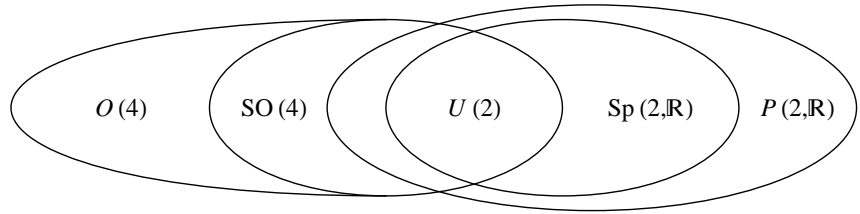}
\end{center}
\caption{Diagram showing the overlap between relevant symmetry groups. The $SO(4)$ symmetry group is the custodial symmetry, $U(2)\cong SO(4)\cap Sp(2,\N{R}) \cong SU(2)_L\times U(1)_Y$ is the global symmetry of the SM, $Sp(2,\N{R})$ is the symmetry group of operators of type $\widehat{C}$, while 
$P(2,\N{R})$ is the symmetry group of operators of type $\widehat{C}^2$. Finally, $O(4)$ is the symmetry group of operators of type $\widehat{B}$. Charge conjugation $C$ will be an element of $P(2,\N{R})^-\cap SO(4)$ where $P(2,\N{R})^-=P(2,\N{R})-Sp(2,\N{R})$, and the same is true for 
the matrix $\mc{J}$ of \eqref{Def:Jk} (with $k=2$). 
\label{figmulti}}
\end{figure*}

The custodial $SO(4)$ symmetry however, is not a subset of $P(2,\N{R})$, and is hence not a symmetry of the operators of the type $\widehat{C}^2$ (nor $\widehat{C}$). See figure \ref{figmulti} for a diagram showing the intersections of the most important symmetry groups in this article.

%%%%%%%%%%%%%%%%%%%%%%%%%%%%%%%%%%%%%%%%%%%%%%%%%%%%%%%%%%%%%%%%%%%%%%%%%

\section{Symmetries of the kinetic terms}
\label{sec:SymmetriesOfTheKineticTerms}
%%%%%%%%%%%%%%%%%%%%%%%%%%%%%%%%%%%%%%%%%%%%%%%%%%%%%%%%%%%%%%%%%%%%%%%%%
We now turn to the (global) symmetries of the 
kinetic terms of the NHDM Lagrangian,
\begin{align}
  K=\sum_{n=1}^N	\left[\left(\partial^\mu + G^\mu \right)\Phi_n(x)\right]^\dag \left[\left(\partial_\mu + G_\mu \right)\Phi_n(x) \right],
\end{align}
with
\begin{align}\label{def:G}
	G^\mu= ig\frac{\sigma^i}{2}W_i^\mu + ig'YB^\mu.
\end{align}
We will let $K_i$ denote the terms of the $i$'th order in the gauge fields.

\subsection{Possible extensions of the global $SU(2)_L \times U(1)_Y$ symmetry}
\label{sec:AnO4SymmetryOfTheKineticTerms}
Now we will start investigating the possibility of having discrete or continuous symmetries beyond the global $SU(2)_L\times U(1)_Y\cong U(2)$ gauge symmetry\footnote{More precisely we have $SO(3) \times U(1) \cong(SU(2)_L\times U(1)_Y)/\N{Z}_2 \cong  U(2)$, where the divisor $\N{Z}_2$ is necessary since multiplication by $-I$ can be expressed both by $SU(2)$ and $U(1)$. We can hence identify elements $(-U)\times U'\sim U \times (-U')$, where $U \in SU(2)_L$ and $U' \in U(1)_Y$. Only multiplication by $-I\in U(2)$ can be expressed both by $U(1)$ and $SU(2)$: Assume $U=\exp(i \alpha)I \in SU(2)$. Then $1=\det(U)=\exp(2\alpha i )$, and hence $\alpha=\pi$, i.e.~$\exp(i \alpha)=-1$. The direct product $SU(2)_L\times U(1)_Y$ is therefore in reality a double cover of $U(2)$.}, in the scalar sector. We do this by writing the NHDM Lagrangian in real form, i.e.~we write the complex Higgs doublets as real quadruplets. In \cite{Olaussen:2010aq} we demonstrated that the kinetic terms
of the NHDM Lagrangian has a symmetry group containing $SU(2) \times U(N)$ (enhanced to $SU(2)\times Sp(N)$ in the limit $g' \to 0$), when we allowed the different Higgs-fields to transform into each other. In the analysis below, we will not consider the possibility of different Higgs-doublets transforming into each other, since these transformations will not be symmetries of the general $C$ invariant NHDM potential, see appendix \ref{sec:HiggsFamilySymmetries}.

Consider the kinetic terms not involving gauge fields, $K_0$, of the Higgs Lagrangian,  \begin{align}\label{E:K0}
	K_0=\sum_{n=1}^N \partial^\mu \Phi_n^T \partial_\mu
 \Phi_n,
\end{align}
where the complex Higgs doublet $(\Phi_n)_2 = \Psi_n + i\Theta_n$ is written in real form, as a quadruplet
from now on denoted
\begin{align}\label{E:PhiReal}
	\Phi_n= \left(\begin{array}{c}\Psi_n\\\Theta_n,
 \end{array}\right).
\end{align}

We see that we can assign the terms $K_0$ a $O(4)$ symmetry by
\begin{align}\label{E:orthogonal2}
	\Phi_n \to O \Phi_n,
\end{align}
with $O\in O(4)$ and $\Phi_m$ given by \eqref{E:PhiReal}. The terms $K_0$
are invariant under the transformation \eqref{E:orthogonal2} since $O^T O =I$ for
$O\in O(4)$. The global symmetry $SU(2)_L \times U(1)_Y \cong U(2)$ will then be embedded into $SO(4)\subset O(4)$ by the map
\begin{align}\label{Def:rho}
\rho (X) =
\begin{pmatrix}
 \text{Re}(X) & -\text{Im}(X) \\
 \text{Im}(X) & \text{Re}(X)
\end{pmatrix},
\end{align}
cf.~appendix B of \cite{Olaussen:2010aq}.
Then the image of $U(2)$ under $\rho$, 
\ba
\rho[U(2)]\subset SO(4), 
\ea
is the global symmetry of the SM, when the fields are written as real quadruplets. The map $\rho$ is an isomorphism onto its image $\rho[U(2)]$, and hence $\rho[U(2)]\cong U(2)$ (i.e.~$\rho[U(2)]$ is a way to write $U(2)$ in real form).
 
We can then, as we showed in \cite{Olaussen:2010aq}, write
\begin{align}\label{E:K1}
	K_1 &=\sum_{n=1}^N \partial^\mu (\Phi_n)_2^\dag G_\mu (\Phi_n)_2 + (\Phi_n)_2^\dag G^{\mu \dag} 
	\partial_\mu (\Phi_n)_2 \nn \\
	  &= \sum_{n=1}^N \partial^\mu \Phi_n^T \mc{T}_\mu \Phi_n + \Phi_n^T(- \mc{T}^{\mu}) 
	\partial_\mu \Phi_n
\end{align}
  where the subscript $2$ in $(\Phi_n)_2$ indicates this is the usual complex Higgs doublet, while $\Phi_n$
  is the four-dimensional real vector \eqref{E:PhiReal}. Furthermore, the $4\times 4$ matrix $\mc{T}^\mu$,
\begin{align}
	\mc{T}^\mu=\rho(G^\mu),
\end{align}
  is given by \cite{Olaussen:2010aq}
\begin{align}\label{E:kinetic1}
	{\cal T^\mu}= \left(\begin{array}{cc}
   gW^\mu_I &-gW^\mu_R-g'\,YB^\mu I_2\\
   gW^\mu_R+g'YB^\mu I_2 &gW_I^\mu 
 \end{array}\right),
\end{align}
 with $W^\mu_R = \sum_{i=1,3} W^\mu_i\,  \frac{1}{2}\sigma^i$ where the sum involves the real Pauli matrices $\sigma^1$ and $\sigma^3$. Moreover, $W^\mu_I = i W^\mu_2 \, \frac{1}{2}\sigma^2$, where $\sigma^2$ is the imaginary Pauli matrix. 
   The matrix $I_2$ is the $2\times 2$ identity matrix. 
 
 We then see that the kinetic terms $K_1$ \emph{apparently} are invariant under the $O(4)$ transformation \eqref{E:orthogonal2} if we let the matrix $\mc{T}^\mu$ transform 
\begin{align}\label{E:mcTO4trafo}
	\mc{T}^\mu \to O \mc{T}^\mu O^T
\end{align}
 simultaneously with \eqref{E:orthogonal2}. As we will see, can the transformation \eqref{E:mcTO4trafo} be described as the adjoint action of $O(4)$ on the Lie algebra $u(2)$ (the latter is represented by the matrix $\mc{T}^\mu$). The SM gauge bosons transform as usual under the global 
 $SU(2)_L \times U(1)_Y \cong U(2) \cong \rho[U(2)]\subset SO(4)$, since for $u\in U(2)$
\begin{align}
	u G^\mu u^\dag = \rho(u) \rho(G^\mu) \rho (u^\dag)= \rho(u) {\cal T^\mu} \rho(u)^T,
\end{align}
 where $u G^\mu u^\dag$ is the way the gauge fields transform (globally) in the SM (they transform under the adjoint representation of $SU(2)_L \times U(1)_Y$) \cite{295711}. We will in
 sec.~\ref{sec:TheCustodialSymmetry} note that in the case $g'\to 0$, the combined transformations \eqref{E:orthogonal2} and \eqref{E:mcTO4trafo} give us the custodial
 $SO(4)$ symmetry.
 
 The only problem is that the transformation \eqref{E:mcTO4trafo} might not be well-defined for choices of $O$ beyond the global gauge group $U(2)$. The transformation is only well-defined when it induces consistent transformations of each of the fields; i.e.~it is well-defined when it makes each of the fields transform in the same manner everywhere in the matrix $\mc{T}$. For instance, we cannot accept that $B^\mu$ transforms as $B^\mu \to B^\mu$ in $\mc{T}^\mu_{14}$, while it transforms as $B^\mu \to -B^\mu$ in $\mc{T}^\mu_{41}$. In section \ref{sec:TheAdjointRepresentationOfO4}
 we will investigate what kinds of transformations $O$ make \eqref{E:mcTO4trafo} well-defined.
 
Third, we consider the kinetic terms quadratic in the gauge fields,
\begin{align}\label{E:K2}
	K_2 &= \sum_{n=1}^N (\Phi_n)_2^\dag G^{\mu \dag}  G_\mu (\Phi_n)_2 \nn \\
	&= -\sum_{n=1}^N \Phi_n^T \mc{T}^2 \Phi_n,
\end{align}
 which obviously are invariant under the $O(4)$ symmetry given by eqs.~\eqref{E:orthogonal2} and \eqref{E:mcTO4trafo}, when the latter transformation is well-defined.
 
\subsection{The gauge field Lagrangian}
\label{sec:TheGaugeFieldTensor}

 In this section we want to show that transformation % \eqref{E:orthogonal2} and 
 \eqref{E:mcTO4trafo} also is a symmetry of the kinetic part of the gauge field Lagrangian. The reason for this is to ensure that symmetries contained in $O(4)$ are not violated by higher order diagrams involving diagrams generated by the (kinetic part of the) gauge field Lagrangian.

 Consider the kinetic terms of the gauge field Lagrangian, formulated as the trace of 
the commutator of two covariant derivatives [as given in \eqref{E:CovD2}], 
\begin{align}\label{E:GFTU0}	\mc{L}_{GB}=-\frac{1}{2}\text{Tr}\left(\left(\frac{i}{g}[D_\mu,D_\nu]\right)^2\right)\bigg|_{g'Y\to g}.
\end{align}
A general relation is 
\begin{align}
	\text{Tr}(\rho(X))= 2 \text{Re}(\text{Tr}(X))=2 \text{Tr}(X),
\end{align}
where the last equality is valid when the trace is real, as it is in \eqref{E:GFTU0}.

Then 
\begin{align}\label{E:commPD0}
	\mc{L}_{GB} &= -\frac{1}{4}\text{Tr}\left(\rho\left(\left(\frac{i}{g}[D_\mu,D_\nu]\right)^2\right)\right)\bigg|_{g'Y\to g} \nn \\
	            &=\frac{1}{4g^2}\text{Tr}\left([\rho(D_\mu),\rho(D_\nu)]^2\right)\Big|_{g'Y\to g} \nn \\
	            &=\frac{1}{4g^2}\text{Tr}\left([\mc{T}_\mu,\mc{T}_\nu]^2\right)\Big|_{g'Y\to g} 
\end{align}
 since $\rho$ preserves matrix multiplication and addition, and since the partial derivatives commute. \eqref{E:commPD0} is invariant under the transformation~\eqref{E:mcTO4trafo}, 
 by the cyclic property of the trace.
 
\subsection{Yukawa couplings}
\label{sec:Yukawa}

  The most general Yukawa couplings of the 
  quark sector are of the form \cite{Branco}
\begin{align}\label{E:YukawaQuarks}
  \mc{L}^{QH} = - \bar{Q}_{L} \left(
            \Delta_{j} \tilde{\Phi}_{j} {p_{R}}  + \Gamma_{ j} 
               \Phi_{j} {n_{R}} \right) 
      + \text{h.c.},  
\end{align}
    where summation over $j=1,2,\ldots,N$ is implicit. The symbol $\bar{Q}_L$ denotes a $1 \times 3$ vector
  consisting of the three left-handed quark doublets, 
\begin{equation}\label{E:Yd} 
  \bar{Q}_{L} = \begin{bmatrix} (u_L^\dag, d_L^\dag) &
  (c_L^\dag, s_L^\dag) & (t_L^\dag, b_L^\dag) 
  \end{bmatrix}\gamma^0,
\end{equation}  
  where $n_R$ and $p_R$ denote (following the notation of
  \cite{Branco}) $3 \times 1$ vectors
  consisting of the right-handed quark fields,
\begin{equation}\label{E:Ys} 
  n_{R}= \begin{bmatrix}
       d_R \\ s_R \\ b_R
       \end{bmatrix}, \quad
  p_{R}= \begin{bmatrix}
       u_R \\ c_R \\ t_R
       \end{bmatrix}, 
\end{equation} 
  and where each of these six quark fields is a four-component Dirac
  spinor. $\Gamma_j$ and $\Delta_j$ are arbitrary, complex $3 \times 3$ matrices. Moreover, the complex doublet $\tilde{\Phi}_{j}$ is defined by
\begin{equation}
  \tilde{\Phi}_{j}= -i(\Phi_{j}^\dag \sigma_2)^T, 
\end{equation}
where ${\Phi}_{j}$ is defined in \eqref{E:SMdoub2}.

  The authors of \cite{Pomarol:1993mu,Haber:1992py,Haber:2010bw} has in the context of the 2HDM shown that  
  the imposition of the custodial symmetry on the Yukawa couplings yields constraints on the coupling matrices $\Gamma_j$ and $\Delta_j$, with consequent mass degeneration of the up- and down-type quarks in the Higgs basis (i.e.~mass degeneration of the quarks in the case of vacuum alignment). Following \cite{Haber:2010bw} we may in the NHDM, after a possible 
  change of scalar field basis\footnote{A change of scalar field basis is a transformation of the scalar fields, $\Phi_m \to \sum_n U_{mn} \Phi_n$, where
  $U\in U(N)$.
  Such a transformation changes the NHDM potential, but not the physics nor the symmetries thereof.} \cite{Nishi:2011gc}, 
  write the custodial symmetric Yukawa terms for the quarks as 
\begin{align}\label{E:Ycs}
   	 \mc{L}^{QH} = - \bar{Q}_{L} \left(\Gamma_{j} 
               \check{\Phi}_j \right)
\begin{pmatrix}
       p_{R} \\ n_{R}
\end{pmatrix} 
      + \text{h.c.}, 
\end{align}
  where $\check{\Phi}_j=(\tilde{\Phi}_j,{\Phi}_j)$ is a $2\times 2$ complex matrix, generalizing the bidoublet $\check{\Phi}$ of \eqref{E:MatrixPhi} to the NHDM. 
  \eqref{E:Ycs} is then invariant under the custodial $SO(4)$ symmetry, by transforming $\bar{Q}_{L}\to \bar{Q}_{L} U_L^\dag$, $ \check{\Phi}_j\to  U_L \check{\Phi}_j U_R^\dag $ and
  $(p_R,n_R)^T \to U_R(p_R,n_R)^T$, where $U_L\in SU(2)_L$ and $U_R\in SU(2)_R$.
  
  By comparing eqs.~\eqref{E:YukawaQuarks} and \eqref{E:Ycs} we see that we need to have 
\begin{align}\label{E:constrYQ}
	\Gamma_j = \Delta_j
\end{align}
  for $j=1,\ldots,N$ to get custodial $SO(4)$ symmetric Yukawa terms for the quarks.
  Then the mass matrices of the up and down-type quarks are given by
\begin{align}\label{E:MMud}
	M_u=\Gamma_j v_j^\ast, \quad M_d=\Gamma_j v_j,
\end{align}
  with a implicit sum over $j$. In the case of vacuum alignment, all VEVs $v_j$ are either real, or all VEVs $v_j$ are imaginary. Then the two mass matrices of \eqref{E:MMud} will yield identical masses
  when bidiagonalized, and hence we have mass degeneration of up- and down-type quarks in the case of vacuum alignment. Therefore, in the case of vacuum alignment, the custodial $SO(4)$ symmetry of the quark-Higgs
  coupling terms must be broken (before spontaneous symmetry breaking), to avoid up-down mass degeneration.
  
  If we do not have vacuum alignment, we only necessarily get identical masses from the bidiagonalization
  of the mass matrices of \eqref{E:MMud} in the case $N=1$. If $N\geq 2$, we will generally not get up-down mass
  degeneration, since the matrices $M_u=\Gamma_j v_j^\ast$ and $M_d=\Gamma_j v_j$ (summed over $j$) will infer different diagonal matrices $D$. Still, the presence of several non-zero Yukawa coupling matrices $\Gamma_j$
  generically will infer flavor changing neutral Yukawa interactions (FCNYI) \cite{Branco}, since e.g.~$M_d=\Gamma_j v_j$ and each $\Gamma_j$
  generally will not be diagonal in the same basis. FCNYI can be avoided by demanding that
  all matrices $\Gamma_i$ and $\Delta_j$ are zero except for (i) two matrices with $i= j$; or (ii) two matrices
  with $i\ne j$. The two scenarios can be enforced by discrete symmetries \cite{Branco}. For the 2HDM, scenario (i) is denoted 2HDM type I, while scenario (ii) is known as 2HDM type II.  The minimal supersymmetric standard model (MSSM) contains a Higgs sector which is a specific variant of the 2HDM type II.
    We see that (i) for the NHDM is consistent with a $SO(4)$ symmetric Yukawa sector, when e.g.~$\Gamma_2=\Delta_2\ne 0$,
    while the other coupling matrices equal zero. But this will still infer up-down mass degeneration, since $M_d = \Gamma_2 v_2$ and $M_u = \Gamma_2 v_2^\ast$ are bidiagonalized to the same diagonal matrix. Scenario (ii) is not consistent with a $SO(4)$ symmetric Yukawa sector, since e.g.~$\Gamma_1\ne 0$ while $\Delta_1=0$ is inconsistent with \eqref{E:constrYQ}. 
    
    On the other hand, if we in the NHDM let $\Gamma_1$ and $\Gamma_2$ be diagonal and $\Gamma_j=0$ for $j>2$, we
    can get custodial symmetric Yukawa couplings of the quarks, realistic quark masses and no FCNYI. The constraints on the Yukawa coupling matrices can be enforced by discrete symmetries. 
    The mass matrices will then be diagonal in the same bases as the coupling matrices, and hence we will have no FCNYI. Moreover, the mass matrices can be bidiagonalized (made real) by 
    diagonal matrices. But then the CKM matrix will become diagonal, contrary to experiment.
    
    Another possibility is the case where the Yukawa coupling matrices are proportional to each other,
    a scenario denoted Yukawa alignment \cite{Pich:2009sp}. Then the Yukawa coupling matrices are of the form
\begin{align}\label{E:propY}
	\Gamma_j= d_j \Gamma_1,
\end{align}
    where each $d_j$ is a complex number, and $d_1=1$. When the Yukawa couplings of the quarks are custodial symmetric,
    the quark mass matrices will become
\begin{align}
	M_u= (d_j v_j^\ast) \Gamma_1, \quad M_d= (d_j v_j) \Gamma_1.
\end{align}
  Then the mass matrices and the coupling matrices will be diagonal in the same bases, since they are all proportional, and we will have no tree-level FCNYI. If all constants $d_j$ are real, we will get up-down mass degeneration, since the mass matrices will differ only by an all-over complex phase. But if some of the numbers $d_j$ have a non-zero imaginary part, 
the mass spectra of $M_u$ and $M_d$ will in general differ. Complex constants $d_j$ may also be a source of $CP$ violation. Unfortunately, the proportionality assumption \eqref{E:propY} is in general not radiatively stable \cite{Ferreira:2010xe}, and we may obtain large, radiatively generated FCNYI \cite{Bijnens:2011gd}.
    
     Mechanisms for natural suppression (i.e.~suppression through an exact or approximate symmetry) 
   of FCNYI (where the FCNYI are suppressed, but non-zero) might still allow for custodial symmetric Higgs-quark couplings, without up-down mass degeneration. For instance,
   BGL-type models \cite{Branco:1996bq} where flavor changing couplings of neutral scalars are fixed by quark masses and elements of the CKM matrix. Here the flavor changing couplings are suppressed by some of the measured, small off-diagonal elements of the CKM matrix. Another way of suppressing FCNYI is by heavy neutral scalars, where the scalars have to be in the TeV range \cite{Branco}. Scalar masses in the TeV range are regarded as unnatural, but is consistent with the recently measured, NHDM-sensitive branching fraction of the rare decay $B_s^0 \to \mu^+ \mu^-$, which was found to be in agreement with the SM expectation \cite{:2012ct}. Some fine-tuning of the Yukawa couplings will still be needed to avoid too much FCNYI mediated by the SM Higgs. 
    
    If the VEVs are not aligned, we may, if the NHDM potential is explicitly $CP$ symmetric, have spontaneous $CP$ violation. Moreover, the custodial $SO(3)_V$ symmetry
       will be spontaneously broken.  
       If the potential is $SO(4)$ symmetric, 5 of 6 $SO(4)$ generators will be spontaneously broken by the non-aligned VEVs. 3 of these 5 broken generators will generate the Higgs ghosts. The two other broken generators will yield a pair of charged, nearly massless pseudo-Goldstone bosons (massless to zero'th order in $g'$) \cite{Olaussen:2010aq}. 
   
Imposing the custodial symmetry on the Yukawa coupling terms of the leptonic sector will lead to constraints analogous to \eqref{E:constrYQ} on the coupling matrices of the leptonic sector.

\section{The adjoint representation of $O(4)$}
\label{sec:TheAdjointRepresentationOfO4}
To decide which transformations beyond $U(2)$ make the transformation \eqref{E:mcTO4trafo}
well-defined, we have to consider the adjoint representation of $O(4)$.
We know the standard model gauge fields transform as the adjoint representation of $U(2)$, which is, written in real form, a sub-representation of the adjoint representation of $O(4)$. For a matrix $d\in O(4)-U(2)$ such that 
\eqref{E:mcTO4trafo} is well-defined for $O=d$, the matrix $d$ generates a discrete symmetry of the kinetic terms of the electroweak Lagrangian. Furthermore, if we could find such a $d \in O(4)-P(2,\N{R})$ (where $P(2,\N{R})$ is the symmetry group of the operators of the type $\widehat{C}^2$, defined in \eqref{E:symGrC^2}, see also figure \ref{figmulti}), we could impose this discrete symmetry on the $C$ invariant NHDM Lagrangian and hence avoid terms proportional to the $O(4)$-violating (but $C$ invariant) operators $\widehat{C}^2$. Then we, because of the discrete symmetry, would have excluded these
$O(4)$-violating terms both from the original potential and during renormalization. 
 
\subsection{Some mathematical preliminaries}
\label{sec:SomeMathematicalProperties} 
 Generally, the adjoint representation of a (matrix) Lie group $G$ is a homomorphism
\begin{align}\label{Def:Ad}
	Ad:G\to GL(\mathfrak{g},\N{R}),
\end{align}
where $\mathfrak{g}$ is the Lie algebra of $G$ and
\begin{align}\label{E:adj.Act}
	Ad(g)(X)=gXg^{-1},
\end{align}
for $g\in G$ and $X\in \mathfrak{g}$. Ad is now a representation of $G$, acting on the vector space $\mathfrak{g}$. The linear transformation $Ad(g)$ on the Lie algebra $\mathfrak{g}$ is called the adjoint action of $g$ on $\mathfrak{g}$. The set of all such linear transformations, 
\begin{align}
	Ad[G]=\{Ad(g)\, : \, g \in G \},
\end{align}
is called the adjoint action of (the Lie group) $G$ on (its Lie algebra)
$\mathfrak{g}$. Moreover, the set $GL(\mathfrak{g},\N{R})$ is the set of all linear, invertible transformations $L$,
on the real vector space $\mathfrak{g}$. If the dimension of $\mathfrak{g}$ is $n$ (i.e.~$\mathfrak{g}$ has $n$ basis vectors)
$GL(\mathfrak{g},\N{R})$ consists of all real, invertible $n \times n$ matrices. In the case $G=SO(4)$, and hence $\mathfrak{g}=so(4)$ (which we will consider in the next section), the dimension of $\mathfrak{g}$ is $n=6$, and therefore $GL(so(4),\N{R})$ will consist of all real, invertible $6 \times 6$ matrices.

Denote the image $Ad[G]$ of the adjoint representation $Ad_G$, where $Ad_G \subset GL(\mathfrak{g},\N{R})$. When G is connected, the kernel of the adjoint representation 
is the center $Z(G)$ of $G$ \cite{Ziller}
\begin{align}\label{E:ZG}
	Z(G)=\{g\in G\, |\, \forall x\in G\, (xg=gx)\}.
\end{align}

  The first isomorphism theorem of group theory then gives us
\begin{align}
	{Ad}_G \cong G/Z(G).
\end{align}
    
In the case of $SO(4)$, which is connected, $Z(SO(4))=\{\pm I \}$. Hence
\begin{align}
	Ad_{SO(4)}={SO(4)}/\{\pm I \}.
\end{align}
Now the group $SO(4)$ is not simple, in contrast to $SO(N)$ for $N=3$ and for $N\geq 5$. Therefore $SO(4)$ can be, modulo its center $\{\pm I \}$, written as a direct product \cite{Cornwell:1985xt}
\begin{align}\label{SOiso}
	{SO(4)}/\{\pm I \}\cong SO(3)\times SO(3).
\end{align}
Hence, we have that
\begin{align}\label{E:AdSO3}
	Ad_{SO(4)} \cong SO(3)\times SO(3),
\end{align}
 which we also will see explicitly below.

\subsection{The effect of the adjoint action}
\label{sec:TheEffectOfTheAdjointAction}
We will now explicitly consider the effect of the adjoint action of $O(4)$ (in fact on the Lie algebra $u(2)$, here parametrized by the gauge boson matrix $\mc{T}^\mu$), 
to see if it permits any symmetries beyond the (global) $U(2)=SU(2)_L\times U(1)_Y$ symmetry of the SM. We regard the real variant of $U(2)$;
embedded in $SO(4)$ by the map $\rho$, and try to see if the adjoint action \eqref{E:mcTO4trafo} can be well-defined for any matrices 
\begin{align}
	O\in O(4)-U(2),
\end{align}
  where $U(2)$ in reality is "shorthand" for $\rho[U(2)]$, the image of $U(2)$ under $\rho$. (We will do similar abbreviations many places in this article, as implied by the context.) The sets $U(2)$ and $\rho[U(2)]$ are the same Lie group, but the latter is expressed by real numbers. 
 The matrix $\mc{T}$ only contains the gauge fields of the SM ($W_1, W_2, W_3, B$), that is, gauge fields corresponding to the Lie algebra $u(2)\subset so(4)$ (here $u(2)$ is written in real form, i.e.~$u(2)$ is here shorthand for $\rho[u(2)]$). Hence the adjoint action \eqref{E:mcTO4trafo} often will demand the extra gauge-fields $X$ and $Y$ of $o(4)-u(2)$, for all gauge fields to transform in consistent manners in $\mc{T}$, when the transformation \eqref{E:mcTO4trafo} is carried out with an $O\in O(4)-U(2)$.
We will in the subsequent search for transformations $O\in O(4)-U(2)$ which do not demand the introduction of the extra gauge-fields of $o(4)-u(2)$ to make the adjoint action \eqref{E:mcTO4trafo} consistent (i.e.~well-defined).

A basis for the Lie algebra $so(4)$ is \cite{Olaussen:2010aq}
\begin{align}
 &J_1 = \left(\begin{array}{rrrr}
   0&1&0&0\\-1&0&0&0\\0&0&0&0\\0&0&0&0
 \end{array}\right),\quad
 J_2 = \left(\begin{array}{rrrr}
   0&0&0&0\\0&0&-1&0\\0&1&0&0\\0&0&0&0
 \end{array}\right), \nonumber\\
 &J_3 = \left(\begin{array}{rrrr}
   0&0&0&0\\0&0&0&-1\\0&0&0&0\\0&1&0&0
 \end{array}\right), \quad
 J_4 =\left(\begin{array}{rrrr}
   0&0&0&0\\0&0&0&0\\0&0&0&1\\0&0&-1&0
 \end{array}\right), \nonumber\\
 &J_5 = \left(\begin{array}{rrrr}
   0&0&0&-1\\0&0&0&0\\0&0&0&0\\1&0&0&0
 \end{array}\right),\quad
 J_6 =\left(\begin{array}{rrrr}
   0&0&-1&0\\0&0&0&0\\1&0&0&0\\0&0&0&0
 \end{array}\right).
\end{align}
Now regard the basis of the Lie algebra $so(4)$ given by
\begin{align}\label{E:basisso4}
  \{ &X_i\}= \nn \\
	\{ &\half (J_2 + J_5), \half (J_1 + J_4), \half (J_6 - J_3), \nn \\
	&\half (J_6 + J_3),\half (J_2 - J_5), \half (J_1 - J_4)\}  ,
\end{align}
 where $X_j=\rho\left((i/2)\sigma^j\right)$, $j=1,\ldots,3$, and $X_4=\rho\left((i/2)I_{2}\right)$, 
 i.e.~the generators $X_1,\ldots,X_4$ are real forms of the generators of the $SU(2)_L\times U(1)_Y$ gauge group. 
 Hence $X_1, \ldots X_4$ corresponds to the SM gauge bosons $W_1, W_2, W_3, B$,
 while $X_5, X_6$ would, if we regarded a full $SO(4)$ gauge symmetry, correspond to two non-SM gauge fields $X$ and $Y$.\footnote{\label{F:su2xsu2}The Lie algebra $so(4)\cong su(2)\oplus su(2)$, and in our notation the matrices $X_1, X_2, X_3$ is a basis for one copy of $su(2)$ while $X_4, X_5, X_6$ is a basis for the other copy of $su(2)$. The matrices $X_5, X_6$ are chosen such that the usual commutator rules for the Lie algebra $su(2)$ are valid, for instance will $[X_5,X_6]=-X_4$, just as $[X_2,X_3]=-X_1$. Moreover, $[su(2)\oplus \{0\},\{0\} \oplus su(2) ]=[\sum_{j=1}^3 t_j X_j, \sum_{j=4}^6 t_j X_j]=0$ for $t_1,\ldots,t_6 \in \N{R}$, as the definition of a direct sum of Lie algebras demands.}
 Then, since $\mc{T}= \rho(G)$ (here we suppress the Lorentz index $\mu$),
\begin{align}
	\mc{T} = \sum_{j=1}^4 w_j X_j,
\end{align}
  where 
\begin{align}
	w_j &= g W_j \quad \text{for} \; j=1,2,3, \quad 
 \text{and} \quad w_4 = 2 g' Y B, 
 \end{align}
 and we also define
\begin{align}  
  w_5 = g_X X,  \quad
  w_6 = g_Y Y.
\end{align}

Since the Lie group $SO(4)$ is compact and connected, exponentiation of its Lie algebra generates the whole group: $\exp[so(4)]=SO(4)$.
Moreover, we can write $SO(4)$ as a product of exponentiated generators ("one-parameter subgroups")
\begin{align}\label{E:UssSO4}
	U=\{e^{t_1X_1} e^{t_2X_2}\cdots e^{t_6X_6} \:|\: \vec{t}\in \N{R}^6 \} = SO(4).
\end{align}
Obviously $U\subset SO(4)$, since all the exponentials are elements of $SO(4)$. The equality $U=SO(4)$ will be demonstrated explicitly below, 
cf.~the discussion following \eqref{E:SO3pry}. 
 
 We will now consider each exponentiation $e^{t_iX_i}$ of the generators $X_i$, and see which effect each of them
 has on the Lie algebra $so(4)$ under the adjoint action [the adjoint action will yield a linear transformation on
 so(4), according to \eqref{Def:Ad}]: 
 
 Let $P_{ij}(\theta)$ denote the $6 \times 6$ matrix with elements
\begin{align}
	p_{ii}&= p_{jj}= \cos (\theta) \nn \\
	p_{ij}&= -p_{ji} =\sin (\theta) \nn \\
	p_{kk}&= 1 \quad k \ne i,j, 	
\end{align}
 with all other elements equaling zero. For instance, $P_{24}(\theta)$ is then
 given by
\begin{align}
	P_{24}(\theta)=\left(
\begin{array}{cccccc}
 1 & 0 & 0 & 0 & 0 & 0 \\
 0 & \cos (\theta ) & 0 & \sin (\theta ) & 0 & 0 \\
 0 & 0 & 1 & 0 & 0 & 0 \\
 0 & -\sin (\theta ) & 0 & \cos (\theta ) & 0 & 0 \\
 0 & 0 & 0 & 0 & 1 & 0 \\
 0 & 0 & 0 & 0 & 0 & 1
\end{array}
\right).
\end{align}
Let $so(4)$ be expressed by the basis $\{X_i\}$ of \eqref{E:basisso4}. Then, for example, the effect of $e^{t_1 X_1} \in SO(4)$
on $so(4)$ by the adjoint action is [see eqs.~\eqref{E:mcTO4trafo} and \eqref{E:adj.Act}],
\begin{align}
	e^{t_1 X_1} \sum_{i=1}^6 w_i X_i e^{-t_1 X_1} &=
	e^{t_1 \half (J_2 + J_5)} \sum_{i=1}^6 w_i X_i e^{- t_1 \half (J_2 + J_5)} \nn \\
	&=E_1 \vec{w} \vec{X},
\end{align}
 where $\vec{w} \vec{X}= \sum_{i=1}^6 w_i X_i$ is a general element in $so(4)$, and where the effect is summarized by the 
 $6\times 6$ matrix $E_1$. Hence $Ad(e^{t_1 X_1})$ maps $\vec{w}$ to $\vec{w'}=E_1\vec{w}$.

 Thus, the effect $E_i$ of each $e^{t_i X_i}$ (no sum over $i$) on $so(4)$ by the adjoint action, can then be calculated 
 by the formula $e^{t_i X_i}\vec{w} \vec{X} e^{-t_i X_i}=E_i \vec{w}$ to be
\begin{align}\label{E:Es}
		E_1 &= P_{23}(t_1), \; E_2 =P_{31}(t_2),\; E_3= P_{12}(t_3) \nn \\
	E_4 &= P_{56}(t_4),\; E_5 =P_{64}(t_5),\; E_6= P_{45}(t_6).
\end{align}
  Moreover, the effect $E_r$ of the reflection 
\begin{align}
	r=\left(
\begin{array}{cccc}
 1 & 0 & 0 & 0 \\
 0 & 1 & 0 & 0 \\
 0 & 0 & 1 & 0 \\
 0 & 0 & 0 & -1
\end{array}
\right),
\end{align}
is then 
\begin{align}\label{E:Er}
	E_r=\left(
\begin{array}{cccccc}
 0 & 0 & 0 & 0 & 1 & 0 \\
 0 & 0 & 0 & 0 & 0 & 1 \\
 0 & 0 & 0 & 1 & 0 & 0 \\
 0 & 0 & 1 & 0 & 0 & 0 \\
 1 & 0 & 0 & 0 & 0 & 0 \\
 0 & 1 & 0 & 0 & 0 & 0
\end{array}
\right).
\end{align}

 The effect of the general element $u = e^{t_1X_1}e^{t_2X_2}\cdots e^{t_6X_6}$ of $ U \subseteq SO(4)$, $U$ introduced in \eqref{E:UssSO4}, is
 then given by 
\begin{align}\label{E:effAdjA}
	u \vec{w} \vec{X}  u^{-1} =\vec{w'} \vec{X} \quad \Leftrightarrow \quad
	\vec{w'} = E \vec{w}
\end{align}
where 
\begin{align}\label{E:Eprod}
	E =E_1 E_2 \cdots E_6. 
\end{align}
  Combining eqs.~\eqref{E:Es} and \eqref{E:Eprod}, we get
\begin{align}\label{E:Em}
	E = 
\left(\begin{array}{cc}
	A(t_1,t_2,t_3) & 0_{3\times 3} \\
	0_{3\times 3} & A(t_4,t_5,t_6)
\end{array}\right),
\end{align}
  where 
\begin{align}\label{E:SO3pry}
	A(x,y,z)= \left(
\begin{array}{ccc}
 c_y c_z & c_y s_z & -s_y \\
 c_z s_x s_y-c_x s_z & c_x c_z+s_x s_y s_z & c_y s_x \\
 c_x c_z s_y+s_x s_z & c_x s_y s_z-c_z s_x & c_x c_y
\end{array}
\right),
\end{align}
  which is just the general element of $SO(3)$, written in the "xyz (pitch-roll-yaw) convention" \cite{Weisstein}.
  Hence the effect $E$ of the general element $u$ of the set $U$ in \eqref{E:UssSO4} is just $SO(3)\times SO(3)$, since
  \eqref{E:Em} yields two independent copies of $SO(3)$.
 Since we already, in \eqref{E:AdSO3}, stated that $Ad_{SO(4)}=SO(3)\times SO(3)$, we know that the parametrization $U$ of \eqref{E:UssSO4} covers the whole of $SO(4)$, at least modulo multiplication by 
 $\pm I$, cf.~\eqref{SOiso}, where $I$ denotes the identity. Moreover, since all exponentials $e^{t_j X_j}$ (no sum over $j$) consists of sines and cosines of angles $t_j/2$ we can express multiplication by minus the $4 \times 4$ identity, $-I_4$, in $U$;
\begin{align}
	e^{(t_j+2\pi) X_j}=-e^{t_j X_j}=-I_4 e^{t_j X_j},
\end{align}
 for any fixed $j\in \{1,\ldots,6\}$.
 Hence $U=\pm U = SO(4)$, since all elements of all equivalence classes in $SO(4)/\{\pm I\}$ are elements of $U$, cf.~eqs.~\eqref{SOiso} and \eqref{E:UssSO4}. 
 
 We do not want to introduce gauge bosons beyond the SM, neither before nor after application of the adjoint action.   
 Hence the parameters $w_5, w_6$ and $w_5', w_6'$ must be zero ($w_5, w_5'$ corresponded to a non-SM gauge-boson $X$ through the generator $X_5$, while $w_6, w_6'$ corresponded to a non-SM gauge boson $Y$ through the generator $X_6$, see \eqref{E:basisso4} and the discussion after). Then we have to demand that some of the elements of $E$ equal zero, namely
\begin{align}\label{E:constrWellDef}
	E_{5i}=E_{6i}=0,\quad i=1,\ldots,4.
\end{align}

 We want to find for which values of the angles $t_1,\ldots,t_6$ the effect of the adjoint action $E$ does not force us to introduce non-SM gauge bosons $X$ and $Y$, corresponding to $w_5=g_X X$ and $w_6 = g_Y Y$. It is then sufficient to consider a matrix 
\begin{align}\label{E:EtildeDef}
	\tilde{E}=E_5 E_6 = E(t_1=\ldots=t_4=0), 
\end{align}
  for if 
\begin{align}\label{E:wEtw}
	\vec{w'}=\tilde{E}\vec{w}, 
\end{align}
  with
\begin{align}
	w_5 = w_6 = w_5' = w_6' =0,
\end{align}
 then the matrix $E_1 \cdots E_4$,
\begin{align}\label{E:E1-E4}
	E_1 \cdots E_4 &= E(t_5=t_6=0) \nn \\ &= \left(\begin{array}{cc}
	A(t_1,t_2,t_3) & 0_{3\times 3} \\
	0_{3\times 3} & A(t_4,0,0)
\end{array}\right),
\end{align}
with 
\begin{align}\label{E:SO3pry00}
	A(t_4,0,0)= \left(
\begin{array}{ccc}
 1 & 0 & 0 \\
 0 & c_{t_4}  & s_{t_4} \\
 0 &- s_{t_4} & c_{t_4}
\end{array}
\right),
\end{align}
  does not introduce non-SM gauge fields when applied on the vector $\vec{w'}$ of \eqref{E:wEtw}, that is,
\begin{align}
	\vec{w''}=E_1\cdots E_4 \vec{w'},
\end{align}
  since $w_5'' = w_6'' =0$.

 Combining the expression for $\tilde{E}$ of \eqref{E:EtildeDef} and the relation $\vec{w'}=\tilde{E}\vec{w}$ for this case ($w_5= w_6=w_5'= w_6'=0$) gives us the equation 
\begin{align}\label{E:ConstrExplicit}
\left(\begin{array}{c}
w_1' \\ w_2' \\ w_3' \\ w_4' \\ 0 \\ 0
\end{array}\right)&=\vec{w'}=\tilde{E}\vec{w} \nn \\ &=
\left(\begin{array}{cc}
	I_{3\times 3} & 0_{3\times 3} \\
	0_{3\times 3} & 
\begin{array}{ccc}
 c_{t_5} c_{t_6} & c_{t_5} s_{t_6} & -s_{t_5} \\
 - s_{t_6} &  c_{t_6} & 0 \\
 c_{t_6} s_{t_5} &  s_{t_5} s_{t_6} &  c_{t_5}
\end{array}
\end{array}\right)
\left(\begin{array}{c}
w_1 \\ w_2 \\ w_3 \\ w_4 \\ 0 \\ 0
\end{array}\right),
\end{align}
We hence see that the only non-trivial constraints are $\tilde{E}_{54}=\tilde{E}_{64}=0$.

  The constraints $\tilde{E}_{54}=\tilde{E}_{64}=0$ read 
\begin{align}\label{E:constraints1}
	- \sin(t_6) &= 0, \nn \\
	\cos(t_6) \sin(t_5) &= 0,
\end{align}
  and hence, 
\begin{align}\label{E:constr_t5}
	\sin(t_5) &=0, \nn \\
	\sin(t_6) &=0. 
\end{align}
  So we have
\begin{align}\label{E:SolConstr}
	t_5,t_6= n\pi, \quad n\in \N{Z},
\end{align}
  not necessarily with the same $n$ for both $t_5$ and $t_6$.  
  
  Now we want to check if the constraints \eqref{E:SolConstr} are consistent with any (discrete) symmetries $d\in SO(4)-P(2,\N{R})$, i.e.~symmetries beyond the symmetry group of the operators of the type $\widehat{C}^2$. 
  The general element of $SO(4)$ can be written 
  \begin{align}
	O= e^{t_1X_1} e^{t_2X_2}e^{t_3X_3} e^{t_4X_4} e^{t_5 X_5}e^{t_6  X_6},
\end{align}
  but it is enough to consider matrices\footnote{\label{F:ud} Let $u \in U(2) \subset P(2,\N{R})$ and $d \in SO(4)-P(2,\N{R})$. Then $ud \notin P(2,\N{R})$: Assume
  the opposite, $ud \in P(2,\N{R})$. Then $u^{-1}(ud)=d \in P(2,\N{R})$, a contradiction. Hence, if $d=e^{t_1X_1}\cdots e^{t_6  X_6} \notin P(2,\N{R})$, then $ud\notin P(2,\N{R})$ for $u=(e^{t_1X_1}\cdots e^{t_4  X_4})^{-1} \in U(2)$. Therefore $d'=ud \notin P(2,\N{R})$, where $d'=e^{t_5  X_5}e^{t_6  X_6}$, and we only have to consider matrices of the same form as $d'$.}
  \begin{align}\label{E:GenOred}
	O= e^{t_5 X_5}e^{t_6  X_6} =\left(
\begin{array}{cccc}
 c_5 c_6 & c_5 s_6 & s_5 s_6 & c_6 s_5 \\
 -c_5 s_6 & c_5 c_6 & -c_6 s_5 & s_5 s_6 \\
 -s_5 s_6 & c_6 s_5 & c_5 c_6 & -c_5 s_6 \\
 -c_6 s_5 & -s_5 s_6 & c_5 s_6 & c_5 c_6
\end{array}
\right),
\end{align}
  where we have defined
\begin{align}
	c_i\equiv \cos(t_i/2), \quad s_i\equiv \sin(t_i/2). 
\end{align}
  We hence find that the matrix $O$ of \eqref{E:GenOred} depends on angles $t_i/2$ for $i=5,6$. Then we have to
   regard the cases where 
\begin{align}\label{E:Constrt5t6}
	t_5,t_6 \in \{0,\pi,2\pi,3\pi \},
\end{align}
  cf.~\eqref{E:SolConstr}.
  We then find that 
\begin{align}\label{E:BchSg}
	\cos(t_5/2)\sin(t_6/2)&= \pm 1 \quad \text{or} \quad \sin(t_5/2)\cos(t_6/2)= \pm 1 \nn \\  \Rightarrow O &\in SO(4)\cap P(2,\N{R})^-,
	\end{align}
	while we otherwise (e.g.~for $t_5=t_6=0$) have
\begin{align}\label{E:BchSgb}	
	O &\in SO(4)\cap Sp(2,\N{R})=U(2).
\end{align}
Eqs.~\eqref{E:BchSg} and \eqref{E:BchSgb} are found 
 by comparing the form of $O$, given by eqs.~\eqref{E:GenOred} and \eqref{E:Constrt5t6}, with respectively the forms \eqref{E:symC2inU2} and \eqref{E:symC2inO2kandX-} given in Appendix \ref{sec:TheFormOfSomeSO4}. 
  The set $P(2,\N{R})^-$ in \eqref{E:BchSg} is a component of the symmetry group of the operators of the form $\widehat{C}^2$, see \eqref{E:symGrC^2}. 
 By considering the actual values of the angles $t_5$ and $t_6$ instead of half the angles, eqs.~\eqref{E:BchSg} and \eqref{E:BchSgb} can equivalently be written
\begin{align}
\cos(t_5)\cos(t_6) &= -1 \Rightarrow O \in SO(4)\cap P(2,\N{R})^-, \label{E:BchSg2} \\
	\cos(t_5)\cos(t_6) &= 1 \Rightarrow O \in SO(4)\cap Sp(2,\N{R})=U(2). \label{E:BchSg3}
\end{align}
  \eqref{E:BchSg2} shows that we get symmetries beyond the gauge group $U(2)$ if and only if
  the field $B_\mu$ changes sign, 
\begin{align}\label{E:B-B}
	B_\mu \to -B_\mu,
\end{align}
   since the parameters $w_4', w_4$ corresponding to the transformation of the gauge field $B_\mu$ are given by $w_4'=E_{44}w_4$, see \eqref{E:ConstrExplicit}, where $E_{44}=\cos(t_5)\cos(t_6)$ \eqref{E:Em}. Moreover, \eqref{E:BchSg3} shows that $B_\mu$ transforms as
   \begin{align}\label{E:B+B}
	B_\mu \to B_\mu,
\end{align}
  under (global) $U(2)\cong SU(2)_L\times U(1)_Y$ transformations, as it should according to the SM.
  
Furthermore, in appendix \ref{sec:TheAdjointActionsOfP2NRCapSO4AreWellDefined}
we show that the symmetry group component the symmetries of 
\eqref{E:BchSg2} belong to, can be written
\ba\label{E:OUC}
   SO(4) \cap P(2,\N{R})^-=U(2)S
\ea
  for any 
\ba  
  S\in SO(4) \cap P(2,\N{R})^-.
\ea
  This is especially true for $S= C\in SO(4) \cap P(2,\N{R})^-$, where $C$ is the charge conjugation operator. We
  implemented $C\in SO(4) \cap P(2,\N{R})^-$ as 
\ba\label{E:matrixC}
C=
\left(\begin{array}{rrrr}
   1&0&0&0\\0&1&0&0\\0&0&-1&0\\0&0&0&-1
 \end{array}\right)
\ea 
 in \cite{Olaussen:2010aq}. This corresponds to
 the transformation $C(\Phi)=\Phi^\dag$ of a complex
 scalar doublet. If we want to include the possibility of a complex phase, i.e.~$C(\Phi)=e^{i \alpha}\Phi^\dag$, where
 $\Phi$ is a complex Higgs doublet,
 we still get $C\in  SO(4) \cap P(2,\N{R})^-$ (this can
 be shown by an argument similar to the one in footnote \ref{F:ud}).
 Hence the 
 arguments below are unaltered by a possible introduction of a phase $\alpha \ne 0$. 
 
 Now, 
 since (by appendix \ref{sec:TheAdjointActionsOfP2NRCapSO4AreWellDefined})
\ba 
 SO(4) \cap P(2,\N{R})^-=U(2)C, 
\ea 
 while
\ba  
 SO(4) \cap Sp(2,\N{R})=U(2)
\ea 
 we get that the Lie group $G$ defined by
\ba\label{E:GSP}
  G=SO(4) \cap P(2,\N{R}),
\ea
can be written
\ba\label{E:UUC}
  G=U(2) \cup U(2)C.
\ea
  Since both $U(2)$ and $U(2)C$ are two sets of symmetries of the kinetic Higgs terms, the Lie group $G$ is a symmetry group of the kinetic Higgs terms. There are by eqs.~\eqref{E:BchSg2} and \eqref{E:BchSg3} no symmetries of the kinetic Higgs terms of the SM beyond the group $G$. This result is also derived in \cite{PhD}, and it proves the sometimes-cited claim that the only symmetry of the kinetic terms, except for gauge transformations, is $CP$. The group $G$ is also the maximal symmetry group of a NHDM Lagrangian, provided that the NHDM potential is complicated enough not to allow any HF symmetries (beyond an overall $U(1)$ transformation). Hence $G$ is the largest symmetry group of a NHDM Lagrangian with a sufficiently complicated potential, e.g.~the most general
 $C$-invariant NHDM Lagrangian, see appendix \ref{sec:HiggsFamilySymmetries}. If we only regard the kinetic terms of the NHDM Lagrangian, the maximal symmetry group will be $SU(N)\times G$, where $SU(N)$ are the Higgs family symmetries. The $U(1)_Y$ hypercharge symmetry is, as we have seen, contained in $G$. 

By \eqref{SOiso} we can write $SO(4)\cong SU(2)_L \times SO(3)$ and by footnote \ref{F:su2xsu2} the three generators of $SU(2)_L$ commutes with the three generators of $SO(3)$ ($\cong SU(2)_R/\N{Z}_2$).
Charge conjugation $C$ will, along with $U(1)_Y$, be a part of $SU(2)_R$. Hence $C$ and $U(1)_Y$ will commute with $SU(2)_L$.
 We can therefore write $G = U(2)\cdot \{I,C\}=SU(2)_L\times (U(1)_Y \cdot\{I,C\})$, where $U(1)_Y\cdot \{I,C\}\cong O(2)$, since $U(1)$ may be mapped onto $SO(2)$ by the map $\rho$ given in \eqref{Def:rho}, while $C$ may be mapped to an arbitrary matrix with determinant $-1$.
Now we get that $G\cong SU(2)\times O(2)$,
 but since both $SU(2)$ and $O(2)$ can express multiplication of the fields by $(-I)$, we divide by $\N{Z}_2$ to avoid a double covering of $G$, where $G$ is given in eqs.~\eqref{E:GSP} and \eqref{E:UUC}. Hence we have
 \ba
   G\cong SO(3)\times O(2).
 \ea
  Since $SO(2)$ is a normal subgroup of $O(2)$ [$gSO(2)g^{-1}\subseteq SO(2)$ for all $g \in O(2)$], and since
  $O(2)\cong U(1)_Y\cdot \{I,C\}\cong SO(2)\cdot \N{Z}_2$, $O(2)$ can also be written as the semidirect product $SO(2) \rtimes \N{Z}_2$. Since $C U(2) C \subseteq U(2)$, $U(2)$ will be a normal subgroup of $G$,
   and we can hence also write 
   \ba
   G\cong U(2) \rtimes \N{Z}_2.
   \ea
 Note that $G$ cannot be written as the direct product of $U(2)$ and $\N{Z}_2$. In fact, 
 \ba\label{E:UH}
  G\ncong U(2)\times H,
\ea
 for any group $H$, since the center of $G$ is finite while the center of $U(2)\times H$ will be infinite,
 see Appendix \ref{sec:TheCenterOfG}. 
  
  However, we find no well-defined $SO(4)$ transformation of the kinetic Higgs terms
  beyond the symmetry group $P(2,\N{R})$ of the operator
  $\widehat{C}^2$. Hence, there is no discrete nor continuous symmetry we can impose on the
  NHDM Lagrangian to exclude the terms of the type $\widehat{C}^2$.
(Again, assuming the potential is complicated enough to exclude the possibility of HF symmetries beyond an overall $U(1)$ transformation. This is the case for the most general, explicitly $C$ invariant NHDM potential, cf.~appendix \ref{sec:HiggsFamilySymmetries}.)
The result may, in the context of the 2HDM, also be derived from the classification of the six classes of possible symmetries that may be imposed on the 2HDM, summarized in \cite{Branco:2011iw}. This summary is based on \cite{Ivanov:2006yq,Ivanov:2007de,Ferreira:2009wh}. Among the listed symmetries there is no symmetry or combination of symmetries which infer the constraint $(\lambda_4-\text{Re}(\lambda_5))=0$. 
Here $(\lambda_4-\text{Re}(\lambda_5))$ is the parameter of the $O(4)$ violating, $C$ respecting term in the 2HDM, which we may write $(\lambda_4-\text{Re}(\lambda_5)) \widehat{C}_{12}^2$, cf.~the 2HDM notation given in Appendix \ref{sec:HiggsFamilySymmetries}.\footnote{This notation is essentially the same as in \cite{Branco:2011iw}.} Hence there is no symmetry that can be imposed on the 2HDM to remove
this term.

\subsubsection{The custodial $SO(4)$ symmetry}
\label{sec:TheCustodialSymmetry}
 
  On the other hand, if we also set $g'=0$ (i.e.~$w_4=w_4'=0$, since these are the parameters corresponding to the generator $X_4$ of the $U(1)_Y$ gauge group, see \eqref{E:basisso4} and the subsequent discussion), we see from eqs.~\eqref{E:ConstrExplicit} and \eqref{E:E1-E4}
  that all symmetries are well-defined, and hence the whole
  $SO(4)$ is a symmetry group of the SM Higgs Lagrangian. This is the custodial $SO(4)$ symmetry from section \ref{sec:custodial}, demonstrated in an alternative manner (for the kinetic terms).
\subsection{The adjoint action of elements of $O(4)^-$}
\label{sec:TheAdjointActionOfElementsOfO4}
  We also want to consider the adjoint action of elements of $O(4)^-$, the orthogonal matrices with determinant $-1$, and see if it
  can yield any (discrete or continuous) symmetries of the kinetic terms. The general effect of the adjoint action of an element of $O(4)^-$ can be written\footnote{$O(n)^-=SO(n)R=RSO(n)$ for any $R\in O(n)^-$: We have $R SO(n)\subset O(n)^-$ since $\det(R)\det(S)=-1$ for $S\in SO(n)$, and $O(n)^-\subset  R SO(n)$ since for $O\in O(n)^-$,
$\det (R^T O)=1$, hence $R^T O \in  SO(n)$, and then $O=R (R^T O)\in  R SO(n) $. The effect under the adjoint representation of an arbitrary element $O\in SO(4)$ can be calculated by $E_1 E_2 \cdots E_6$, cf.~eqs.~\eqref{E:effAdjA} and \eqref{E:Eprod}. Then the effect of an arbitrary element $R O\in O(4)^-$ can be calculated by $E_r E_1 E_2 \cdots E_6$, cf.~\eqref{E:Er}.}
\begin{align}\label{E:Eprod2}
	E^- =E_r \cdot E_1 E_2 \cdots E_6,
\end{align}
 cf.~\eqref{E:Eprod}, where $E_r$ is given in \eqref{E:Er}, and we calculate $E^-$ to be of the form
\begin{align}\label{E:adjact-}
	E^- =\left(\begin{array}{cc}
	0_3 & D(t_4,t_5,t_6) \\
	C(t_1,t_2,t_3) & 0_3
\end{array}\right),
\end{align}
  where $0_3$ is the $3 \times 3$ zero matrix, and where the exact expressions of the $3 \times 3$ matrices $C$ and $D$ are irrelevant for 
  the argument below.
  
   In case $E^-$ represented a well-defined adjoint action (i.e.~the non-SM parameters equal zero: $w_5'=w_6'=w_5=w_6=0$), $E^-$ would have the effect
\begin{align}\label{E:ConstrExplicitE-}
\left(\begin{array}{c}
w_1' \\ w_2' \\ w_3' \\ w_4' \\ 0 \\ 0
\end{array}\right)= E^- \left(\begin{array}{c}
w_1 \\ w_2 \\ w_3 \\ w_4 \\ 0 \\ 0
\end{array}\right)=\left(\begin{array}{c}
E^-_{14} w_4 \\ E^-_{24} w_4 \\ E^-_{34} w_4 \\ \sum_{j=1}^3 E^-_{4j} w_j \\ \sum_{j=1}^3 E^-_{5j} w_j \\ \sum_{j=1}^3 E^-_{6j} w_j
\end{array}\right). 
 \end{align}
   But then all gauge fields $W_j, j=1,2,3$ (the Lorentz index $\mu$ is suppressed) are transformed into some real number times the field $B$, since $w_j=g W_j$ for $j=1,\ldots, 3$ and $w_4=2g'Y B$. But this transformation has no inverse transforming the $B$'s ($w_4$) back into the different $W_j$'s ($w_j, j=1,2,3$), since it would have transformed different instances of the field $B$ in a different manner. Hence no such well-defined %adjoint $O(4)^-$ action
    $E^-$, as it appears in \eqref{E:ConstrExplicitE-}, exists, because $E^-$ should have an inverse (an adjoint action is a representation of a group, and all elements of a group has an inverse).   
    Therefore $O(4)^-$ does not provide any new symmetries of the kinetic Higgs terms of the SM (or NHDM) Lagrangian. 
    
     The component $O(4)^-$ of $O(4)$ does not provide any new symmetries in the limit $g'\to 0$ (i.e.~$w_4=w_4'=0$) either: In this case $W_j, j=1,2,3$ would have to be mapped to $0$, so this mapping could have no inverse, hence it cannot be an element of (the adjoint representation of) $O(4)$. This means the custodial
 $SO(4)$ symmetry cannot be extended by any (discrete or continuous) symmetry in $O(4)$. The result is also derived in \cite{PhD}. It is also proven by one of the results in \cite{Pilaftsis:2011ed}. In this article all 13 possible accidental symmetry groups of a 2HDM potential, together with the kinetic Higgs terms in the limit $g' \to 0$, are determined. Here the maximal symmetry group, in case of a potential consisting of all possible $O(4)$ symmetric terms of the (in our terminology) type $\widehat{B}$ and $\widehat{B}^2$ (but transformed into a specific basis, see appendix \ref{sec:HiggsFamilySymmetries}), is determined to be $SO(3)\times SU(2)_L \cong SO(4)$. I.e.~the custodial $SO(4)$ symmetry is the maximal symmetry group of the kinetic Higgs terms in the limit $g'\to 0$ (when we are not regarding $SU(N)$ HF-symmetries).

\section{Summary}
\label{sec:Summary}

By studying which $O(4)$ transformations 
$\mc{T}^\mu \to O^T \mc{T}^\mu O$ [cf.~\eqref{E:mcTO4trafo}] of the gauge bosons
are well-defined, we found the maximal set of symmetry transformations of the kinetic
Higgs terms in the SM to constitute the Lie group $G=SO(4)\cap P(2,\N{R}) \cong SO(3)\times O(2)$. The maximal symmetry group of the kinetic terms of the NHDM was then $SU(N)\times G$. The Lie group
$P(2,\N{R})$ was the symmetry group of operators of the form $\widehat{C}^2$, that is the $O(4)$ violating terms in the general, explicitly $C$ (i.e.~$CP$, cf.~comment after \eqref{E:CovD2}) invariant NHDM potential. Hence, we could find no discrete nor continuous symmetry that is a symmetry of the kinetic terms and the quadratic (in the Higgs fields) terms of the type $\widehat{B}$, while not a symmetry of the terms of the type $\widehat{C}^2$. To show this, we also used the fact that the Higgs family symmetries of a potential containing all terms $\mu_{mn} \widehat{B}_{mn}$, $m\leq n$, also are symmetries of the terms of the type $\widehat{C}^2$, cf.~appendix \ref{sec:HiggsFamilySymmetries}. This means there 
is no symmetry we can impose on the general NHDM Lagrangian, to exclude the $O(4)$ violating terms from the NHDM potential, and to prevent them occurring as counterterms during coupling constant renormalization.

As implied above, we found there are no symmetries of the kinetic
Higgs terms in the negative determinant component $O(4)^-$ of $O(4)$.
This was so, even in the limit $g'\to 0$. Hence, the custodial $SO(4)$ symmetry cannot be extended by elements of $O(4)^-$.

In sec.~\ref{sec:Yukawa} we saw that if we impose the $SO(4)$ custodial symmetry on the Yukawa couplings of the quarks, it will infer
up-down mass degeneration, if the VEVs are aligned (e.g.~by all VEVs being real). If the VEVs are not aligned, the up- and down-type mass matrices $M_u$ and $M_d$ in general will yield distinct mass spectra. Then we might obtain models with custodial symmetric Yukawa couplings of the quarks, realistic quark masses and CKM matrix, and suppressed flavor changing neutral Yukawa interactions.
If the NHDM potential is $SO(4)$ symmetric (before spontaneous symmetry breaking), the aforementioned non-alignment of the VEVs also will cause a pair of light, charged pseudo-Goldstone bosons to emerge. 

\acknowledgments

{The author wants to thank Kåre Olaussen, Per Osland, 
David Fleming and Michael Kachelrie\ss $\,$ for helpful conversations.}

\appendix
\section{Higgs family symmetries}
\label{sec:HiggsFamilySymmetries}
 In this appendix we will show that the $U(N)$ Higgs family (HF) symmetries\footnote{Here we also consider the $U(1)$ factor [of $U(N)$] as a HF symmetry, in addition to the $SU(N)$ HF symmetries. Overall multiplication by an $U(1)$ factor is in fact already taken care of by $U(1)_Y$, which we usually consider as a gauge group symmetry and not a HF symmetry.} \cite{Ferreira:2008zy} of a potential containing all terms $\mu_{mn}\widehat{B}_{mn}$, $m\leq n$,
 also are Higgs family symmetries of the bilinears of the type $\widehat{C}_{mn}$. Hence there is no
 HF symmetry we can impose on the general, explicitly $C$ invariant NHDM potential, which exactly excludes the $O(4)$ violating
 terms, i.e.~the terms of the form $\widehat{C}^2$.

 Under a HF transformation the bilinear $\widehat{B}_{kk}\equiv \Phi_k^\dag \Phi_k$ 
 is transformed in the following manner
 \ba
    \Phi_k^\dag \Phi_k \to &\sum_{m,n=1}^N (U_{km} \Phi_m)^\dag (U_{kn} \Phi_n) \nn \\ &= \sum_{m,n=1}^N U_{km}^\ast U_{kn} \Phi_m^\dag \Phi_n.
 \ea
 If this $U(N)$ transformation (i.e.~HF symmetry) keeps $\Phi_k^\dag \Phi_k$ invariant, then (no sum over $k$)
 \ba\label{E:HFBkkUU}
 U_{km}^\ast U_{kn} = 1,& \quad \text{when} \; m=n=k, \nn \\
 U_{km}^\ast U_{kn} = 0,& \quad \text{otherwise}.
 \ea
 This is so because the set $\{\Phi_m^\dag \Phi_n \}_{m,n\leq N}$ (with $N^2$ elements) is linearly independent: The set $\{\Phi^\dag_n, \Phi_m\}$ can be regarded classically as a set of $2N$ independent
 functions. Assume that
 \ba
 \sum_{i,j=1}^N a_{ij}\Phi^\dag_i \Phi_j\equiv 0. 
 \ea
 If we then differentiate by $\partial^2/(\partial \Phi_m^\dag \partial \Phi_n)$, for arbitrary $m$ and $n$, we get that
 $a_{mn}=0$, i.e.
 \ba
 a_{mn}=0 \quad \text{for all}\quad m,n. 
 \ea
 This again means that the set $\{\Phi_m^\dag \Phi_n \}_{m,n\leq N}$ is linearly independent.
 
 \eqref{E:HFBkkUU} then gives us that
 the most general HF transformation which keeps a term $\mu_{kk}\widehat{B}_{kk}$ (no sum over $k$) invariant, $1 \leq k \leq N$, is
 \ba\label{E:HFBkk}
    \Phi_k \to \exp(i\alpha_k) \Phi_k.
 \ea
 When we want to
 keep the term $\mu_{kk}\widehat{B}_{kk}$ invariant for all $k$, each bilinear has to transform as given in
 \eqref{E:HFBkk}, with a possible different phase $\alpha_k$ for each $k$.

 Now, if the potential we are regarding also contains all terms of the type $\mu_{mn}\widehat{B}_{mn}$ for $m< n$, in addition to the terms $\mu_{kk}\widehat{B}_{kk}$,  there will be further restrictions on the allowed HF symmetries. 
 The bilinear $\widehat{B}_{mn}$, $m<n$, now transforms as
 \begin{align}\label{E:trafoB}
    \widehat{B}_{mn} &\equiv \half(\Phi_m^\dag \Phi_n+\Phi_n^\dag \Phi_m)\nn \\
    &\to \half(U_{mm}^\ast \Phi_m^\dag U_{nn}\Phi_n+U_{nn}^\ast \Phi_n^\dag U_{mm}\Phi_m) 
 \end{align}
 (no sum over $m,n$) by eq.~\eqref{E:HFBkkUU} [valid for all $k$]. When
 $\widehat{B}_{mn}$, $m<n$, is kept invariant, that is,
 \begin{align}
    &U_{mm}^\ast \Phi_m^\dag U_{nn}\Phi_n+U_{nn}^\ast \Phi_n^\dag U_{mm}\Phi_m \nn \\ &-\Phi_m^\dag \Phi_n-\Phi_n^\dag \Phi_m=0,
 \end{align}
 (no sum over $m,n$), we must have
 \ba\label{E:UmmUnn}
    U_{mm}^\ast U_{nn}=1,
 \ea
 since the two terms $\Phi_m^\dag \Phi_n$ and $\Phi_n^\dag \Phi_m$ are linearly independent. 
 \eqref{E:UmmUnn} infers that the corresponding angles are identical,
 \ba
    \alpha_m=\alpha_n,
 \ea
 where $U_{kk}=\exp(i \alpha_k)$.
 Hence the HF symmetries that keep all terms $\mu_{mn}\widehat{B}_{mn}$ simultaneously invariant for 
 all $m,n$ with $1\leq m \leq n \leq N$, is an $U(1)$ transformation 
 \ba\label{E:oaU(1)}
 \Phi_j \to \Phi_j \exp(i\alpha),
 \ea 
 for all $j$, $1\leq j \leq N$. 
 But this transformation also leave all terms $\widehat{C}_{mn}$ and $\widehat{C}_{mn} \widehat{C}_{m'n'}$
 invariant, and hence the transformation cannot be applied to remove terms of the type $\widehat{C}^2$. 
  
The result in \eqref{E:oaU(1)} and above is consistent with a result in \cite{Pilaftsis:2011ed}. Here the maximal symmetry group of the general $C$ invariant 2HDM potential (expressed in the diagonally reduced basis)  
is found to be the group generated by $CP$ (i.e.~$C$) and $U(1)_Y$.

The general 2HDM potential of \cite{Pilaftsis:2011ed} can be expressed by the bilinear operators $\widehat{B}$ and $\widehat{C}$,  
in terms of the parameters given in \cite{Pilaftsis:2011ed} as
\begin{align}\label{E:pot2HDM}
 V &= -\mu_1^{2}\widehat{B}_{11} - \mu_2^{2}\widehat{B}_{22}
 -2 \text{Re}(m^2_{12})\widehat{B}_{12} +2 \text{Im}(m^2_{12}) \widehat{C}_{12}
 \nn \\
 &+ \lambda_{1} \widehat{B}_{11}^2 + \lambda_{2} \widehat{B}_{22}^2 
 +\lambda_{3} \widehat{B}_{11}\widehat{B}_{22} 
 + \left[\lambda_{4}+\text{Re}(\lambda_{5})\right] \widehat{B}_{12}^2 
 \nn \\&+ 2\text{Re}(\lambda_{6}) \widehat{B}_{11} \widehat{B}_{12}
 + 2\text{Re}(\lambda_{7}) \widehat{B}_{22} \widehat{B}_{12}
 \nn \\&-2\text{Im}(\lambda_{5}) \widehat{B}_{12} \widehat{C}_{12} 
 -2\text{Im}(\lambda_{6}) \widehat{B}_{11} \widehat{C}_{12}
 \nn \\
 &-2\text{Im}(\lambda_{7}) \widehat{B}_{22} \widehat{C}_{12}
 +\left[\lambda_{4}-\text{Re}(\lambda_{5})\right] \widehat{C}_{12}^2.
\end{align}
All parameters in \eqref{E:pot2HDM} are real (in some cases by taking the real or imaginary part of a complex parameter).
In the diagonally reduced basis, we can furthermore set $\text{Im}(\lambda_{5})=0$ and $\lambda_{6}=\lambda_{7}$.
 Then the potential symmetric under $C$ and $U(1)_Y$ (in addition to $SU(2)_L$) corresponds to the constraints
 $\text{Im}(m^2_{12})=0$ and $\text{Im}(\lambda_{6})=\text{Im}(\lambda_{7})=0$, i.e.~we are left with the potential 
 \begin{align}\label{E:pot2HDMCinv}
 V &= -\mu_1^{2}\widehat{B}_{11} - \mu_2^{2}\widehat{B}_{22}
 -2 \text{Re}(m^2_{12})\widehat{B}_{12} \nn \\
 &+ \lambda_{1} \widehat{B}_{11}^2 + \lambda_{2} \widehat{B}_{22}^2 
+ \lambda_{3} \widehat{B}_{11}\widehat{B}_{22} \nonumber \\ 
 &+ \left[\lambda_{4}+\text{Re}(\lambda_{5})\right] \widehat{B}_{12}^2 
 + 2\text{Re}(\lambda_{6}) (\widehat{B}_{11} \widehat{B}_{12} 
 +  \widehat{B}_{22} \widehat{B}_{12}) \nn \\
 &+\left[\lambda_{4}-\text{Re}(\lambda_{5})\right] \widehat{C}_{12}^2,
\end{align}
 which is the most general $C$ invariant 2HDM potential, expressed in the diagonally reduced basis.  We note that it, according to both \cite{Pilaftsis:2011ed} and \eqref{E:oaU(1)}, has no HF symmetries (beyond $U(1)_Y$, when we consider this as a HF symmetry). 
 
 In the case $\left[\lambda_{4}-\text{Re}(\lambda_{5})\right]=0$, we see that the symmetry is enhanced, since the potential then only contains $O(4)$ symmetric operators (of the type $\widehat{B}$). When we include the 
 Kinetic Higgs terms in the limit $g' \to 0$, the maximal total symmetry becomes $SO(4)$, cf.~sec.~\ref{sec:TheAdjointActionOfElementsOfO4}. We also noted in sec.~\ref{sec:TheAdjointActionOfElementsOfO4} that this is
 consistent with a result in \cite{Pilaftsis:2011ed}, which determines this 2HDM to have the maximal symmetry
 group $SO(3)\times SU(2)_L \cong (SU(2)\times SU(2)_L)/\N{Z}_2 \cong SO(4)$. Finally we note that the greatest
 symmetry group of any 2HDM potential, with kinetic terms in the limit $g'\to 0$, is in \cite{Pilaftsis:2011ed} calculated to be
 $SO(5)\times SU(2)_L$. This symmetry corresponds to a potential with the additional restrictions
 $\mu_1^{2}=\mu_2^{2}$, $m^2_{12}=0$, $\lambda_{1}=\lambda_{2}$, $\lambda_3= 2\lambda_1$ and $\lambda_{j}=0$ for
 $j=4,5,6$ and $7$.
 This is in agreement with \cite{Olaussen:2010aq}, where we determined the symmetry group
 of the kinetic terms in the limit $g'\to 0$ to be $Sp(N)\times SU(2)_L$, where $Sp(N)$ is the quaternionic 
 symplectic group. The agreement is proven by the isomorphism $SO(5)\cong Sp(2)/\N{Z}_2$ (see \cite{Cornwell:1985xt} p.~430), where the $\N{Z}_2$   
 factor reflect that we can express multiplication of the fields by $-I$ both in $SU(2)_L$ and $Sp(2)$ (since $SU(2)_R\subseteq Sp(N)$).
  
\section{The form of some $O(2k)$ matrices}
\label{sec:TheFormOfSomeSO4}
Let $S\in O(2k)$ (the case $k=2$ is the most interesting for this article), and let
\begin{align}\label{Def:Jk2}
	 {\cal J}_k = {\cal J}=\left(\begin{array}{rr}0_k&I_k\\-I_k&0_k\end{array}\right).
\end{align}
   Then the condition $S^T {\cal J} S= {\cal J}$ [i.e.~$S\in Sp(k,\N{R})$]
can be written 
\begin{align}\label{E:SinU2}
	{\cal J}S &= (S^T)^{-1} {\cal J}, \nn \\
	{\cal J}S &= S {\cal J}, 
\end{align}
  which, by the definition \eqref{Def:Jk2} of ${\cal J}$, forces the solutions of \eqref{E:SinU2} to be exactly the matrices in $O(2k)$ of the form
\begin{equation}\label{E:symC2inU2}
 S = \left(\begin{array}{cc}A&B\\
   -B& A \end{array}\right),
\end{equation}
for arbitrary $k\times k$ matrices $A, B$.

Again, let $S\in O(2k)$. Then the equation the condition $S^T {\cal J} S = - {\cal J}$
[which is the definition of $S\in P(k,\N{R})^-$], can be written
\begin{align}
	{\cal J}S    = -S {\cal J}, 
\end{align}
  which forces the matrices $S$ 
to be exactly the ones in $O(2k)$ of the form
\begin{equation}\label{E:symC2inO2kandX-}
 S = \left(\begin{array}{cc}A&B\\
   B&-A \end{array}\right),
\end{equation}
for arbitrary $k\times k$ matrices $A, B$.  

\section{The component $SO(4) \cap P(2,\N{R})^-$}
\label{sec:TheAdjointActionsOfP2NRCapSO4AreWellDefined}
 In this appendix we want to prove that
\begin{align}\label{E:Ud=}
	U(2) S = SO(4) \cap P(2,\N{R})^-,
\end{align}
 for any $S \in SO(4)\cap P(2,\N{R})^-$.
 
First, we claim the set $P(2,\N{R})^-$ is given by
\begin{align}\label{E:P-Sp}
	P(2,\N{R})^-=Sp(2,\N{R})\,S =S \, Sp(2,\N{R}),
\end{align}
  for any $S\in P(2,\N{R})^-$:
  
  $\mathbf{P(2,\N{R})^- \subseteq Sp(2,\N{R}) S}$:
  Let $S' \in P(2,\N{R})^-$. Then $S' S \in Sp(2,\N{R})$ since
\begin{align}\label{E:SS}
	(S'S)^T\mc{J}(S'S)= S^T(-\mc{J})S=\mc{J}.
\end{align}
  Then $S' = T S$ for $T=S'S^{-1} \in Sp(2,\N{R})$, i.e.~$S'\in Sp(2,\N{R}) S$. This is true since $S',S^{-1}\in P(2,\N{R})^-$ infer $S'S^{-1} \in Sp(2,\N{R})$ by \eqref{E:SS}.
  ($S^{-1}\in P(2,\N{R})^-$ since $S \in  P(2,\N{R})^-$.)
  Similarly with $S\, Sp(2,\N{R})$.
  
$\mathbf{P(2,\N{R})^- \supseteq Sp(2,\N{R}) S}$: On the other hand, if $T \in Sp(2,\N{R})$, then 
\begin{align}
	(T S)^T\mc{J}(T S)=S^T \mc{J} S =-\mc{J},
\end{align}
 so then $T S \in P(2,\N{R})^-$. Similarly, $S T \in P(2,\N{R})^-$.  
  
  Now we can derive \eqref{E:Ud=}: Let $S \in P(2,\N{R})^-\cap SO(4)$. 
   Then
\begin{align}
	U(2)S&= (SO(4)\cap Sp(2,\N{R}))S \nn \\
	&=(SO(4)S)\cap (Sp(2,\N{R})S)
     \nn \\	
	&=SO(4)\cap P(2,\N{R})^-,
\end{align}
 the last equality by \eqref{E:P-Sp}.

\section{The center of $G$}
\label{sec:TheCenterOfG}
In this appendix we will show that the Lie group $G=P(2,\N{R})\cap SO(4)$ has a finite center $Z(G)=\{\pm I_4 \}\cong \N{Z}_2$. Hence $G$ cannot be isomorphic to a group with a infinite center, especially not groups
of the form $U(2)\times H$, $H$ arbitrary. 

First, $G$ is a Lie group since the intersection of two (topologically) closed
subsets of $GL(n,\N{R})$ is a (topologically) closed subset of $GL(n,\N{R})$, and hence a Lie group.
The center $Z(G)$ of $G$ consists of the elements in $G$ which commute with all elements in $G$, cf.~\eqref{E:ZG}.
Let $Y\in Z(G) \subset G$. Then
\ba\label{E:YinP}
  Y^T\mc{J}Y=\pm \mc{J},
\ea
since $Y\in P(2,\N{R})$. Moreover, since $Y$ commutes with any element in $G$, $Y^T Y\mc{J}=\pm \mc{J}$,
and hence
\ba
  Y^TY=\pm I.
\ea
Since $Y \in G \subset SO(4)$ we get $I=\pm I$, and then + is the right sign in \eqref{E:YinP},
\ba\label{E:YinSp}
  Y^T\mc{J}Y= + \mc{J},
\ea
which again infer that 
$Y\in Sp(2,\N{R})$.  

This means 
\ba
Y\in Sp(2,\N{R})\cap SO(4)=U(2),
\ea
and thus $Y$ is an element of $U(2)$ which commutes with all elements of $G\supset U(2)$, and hence we
also have
\ba
Y\in Z(U(2)).
\ea

Now $Z(U(2))\cong U(1)$, or more concretely  
\ba
  Z(U(2))=\{ \lambda I_2 : \lambda \in \N{C} \wedge |\lambda|=1 \},
\ea
which is a consequence of Schur's lemma.
Expressed as a $4 \times 4$ real matrix by the map $\rho$, an element $Y\in Z(U(2))$ is of the form
\ba
   Y =\left(
\begin{array}{cc}
	\cos \theta I_2 & -\sin \theta I_2  \\
	\sin \theta I_2  & \cos \theta I_2
\end{array} \right),
\ea
where $\cos \theta = \text{Re}(\lambda)$ and $\sin \theta = \text{Im}(\lambda)$.

Since $Y$ also is an element in $Z(G)$, it also needs to commute with all elements of $G$, especially
$C\in P(2,\N{R})^- \cap SO(4) \subset G$. This means $YC=CY$, that is,
\begin{align} 
&\left(
 \begin{array}{cc}
	c_\theta I_2 & -s_\theta I_2  \\
	s_\theta I_2  & c_\theta I_2
\end{array} \right) 
\left(
\begin{array}{cc}
	I_2 & 0_2  \\
	0_2  & -I_2
\end{array} \right) \nn \\
&=
\left(
\begin{array}{cc}
	I_2 & 0_2  \\
	0_2  & -I_2
\end{array} \right) \left(
\begin{array}{cc}
	c_\theta I_2 & -s_\theta I_2  \\
	s_\theta I_2  & c_\theta I_2
\end{array} \right)
\end{align}
which yields $s_\theta \equiv\sin \theta= 0$ and $c_\theta \equiv \cos \theta=\pm 1$, which again give us
\ba
   Y=\left(
\begin{array}{cc}
\pm	I_2 & 0_2  \\
	0_2  & \pm I_2
\end{array} \right)= \pm 
I_4. 
\ea 
 Hence we have derived the result $Z(G)\subset \{\pm I_4\}$, and since both $\pm I_4$ evidently
 commute with all of $G$, we get
 \ba
  Z(G)= \{\pm I_4\} \cong \N{Z}_2,
 \ea
  i.e.~a finite center.
  
  On the other hand,
\begin{align}
	Z(U(2)\times H) &= Z(U(2))\times Z(H) \nn \\&\cong U(1) \times Z(H),
\end{align}  
  which is infinite, since $U(1)$ is infinite. Thus we have $G\ncong U(2)\times H$ for any group $H$.

%\section*{References}

\end{document}